\documentclass[a4paper,12pt]{article}

\pdfoutput=1 

\usepackage[hmargin=.71in,vmargin=1.1in]{geometry}
\usepackage{indentfirst}
\usepackage{amsfonts}
\usepackage{mathrsfs}
\usepackage{amsmath}
\usepackage{amssymb}
\usepackage{authblk}
\usepackage{cite}
\usepackage{xcolor}
\usepackage{mathtools}
\usepackage{tensor}
\usepackage{physics}
\usepackage{graphicx}
\usepackage{bm}
\usepackage{upgreek}
\usepackage{braket}
\usepackage{color,soul}
\usepackage{csquotes}
\usepackage{caption}
\usepackage{subcaption}
\usepackage{slashed}
\usepackage{mathtools}

\usepackage[bookmarksnumbered=true,bookmarksopen=true]{hyperref}
 \hypersetup{colorlinks,
             linkcolor=[rgb]{0,0.3,0.6}, 
             citecolor=[rgb]{0,0.3,0.6}, 
             urlcolor=[rgb]{0,0.3,0.6}}

\newcommand\e{\varepsilon}
\newcommand\mi{\mathrm{i}}
\newcommand\me{\mathrm{e}}

\newcommand{\dif}{\mathrm{d}}

\DeclareMathOperator{\diag}{diag}

\DeclareMathOperator{\arctanh}{arctanh}

\usepackage{tikz}
\usetikzlibrary{matrix}

\begin{document}

\title{\Large\textbf{Pseudo-Hermitian Chebyshev differential matrix and non-Hermitian Liouville quantum mechanics}}

\author{Chen Lan\thanks{stlanchen@126.com}}
\author{Wei Li}
\author{Huifang Geng\thanks{genghf@ytu.edu.cn}}

\affil{\normalsize{\em Department of Physics, Yantai University, 30 Qingquan Road, Yantai 264005, China}}

\date{ }

\maketitle

\begin{abstract}
The spectral collocation method (SCM) exhibits a clear superiority in solving ordinary and partial differential equations compared to conventional techniques, 
such as finite difference and finite element methods. 
This makes SCM a powerful tool for addressing the Schr\"odinger-like equations with boundary conditions in physics. 
However, the Chebyshev differential matrix (CDM), commonly used in SCM to replace the differential operator, 
is not Hermitian but pseudo-Hermitian. 
This non-Hermiticity subtly affects the pseudospectra and 
leads to a loss of completeness in the eigenstates.
Consequently, several issues arise with these eigenstates. 
In this paper, we revisit the non-Hermitian Liouville quantum mechanics by emphasizing the pseudo-Hermiticity of the CDM 
and explore its expanded models.
Furthermore, we demonstrate that the spectral instability can be influenced by the compactification parameter.

\end{abstract}

\tableofcontents

\section{Introduction}
\label{sec:intr}

PT-symmetric quantum mechanics has demonstrated significant vitality since its inception 
and has found widespread application in several fields of physics, 
as highlighted in the recent review by its founders  \cite{Bender:2023cem}. 
Here, P and T denote parity reflection and time reversal, respectively. 
This quantum theory asserts that if a non-Hermitian Hamiltonian satisfies parity and time-reversal symmetries, 
its eigenvalues will be real \cite{Bender:1998ke,Bender:2017hqr}. 
Furthermore, if the corresponding complexified classical trajectory is closed, 
the eigenvalues will be discrete \cite{Bender:1998gh,Bender:2014bna}. 
To restore the completeness of eigenstates, the metric operator \cite{Mostafazadeh:2002pht} is introduced in the Hilbert space. 
This operator facilitates a similarity transformation that allows one to 
find a Hermitian equivalent Hamiltonian \cite{Mostafazadeh:2005wm,Jones:2006qs,Nanayakkara:2012kc}, 
which is isospectral with the original non-Hermitian Hamiltonian. 
Complexified Liouville quantum mechanics, as one of the PT-symmetric models, has garnered significant interest \cite{Cannata:1998bp,Curtright:2005zk,Bender:2014bna,Bagrets:2016cdf}, 
not only because it is completely integrable and involves non-perturbative aspects, 
but also due to its various applications in condensed matter physics \cite{Bagrets:2016cdf}, string theory  \cite{Nakayama:2004vk,Li:2019mwb,Mertens:2020hbs,Betzios:2020nry}, 
and quantum cosmology \cite{Andrianov:2016ffj,Andrianov:2018wdx,Lan:2021vha}, among others.

Among the various techniques for handling non-normal operators, 
such as the spectral parameter power series method \cite{Kravchenko:2008rep,Kravchenko:2010spe,Rabinovich:2019spe,Barrera:2019num,Barrera:2022pow}, 
the {\em spectral collocation method} (SCM) \cite{Trefethen:2020sp,Trefethen:2000smt} stands out 
as a robust tool for analyzing spectral properties and instabilities. 
By leveraging its connection to the pseudospectrum, 
SCM provides an effective framework for examining the spectra and stability characteristics of non-normal operators.
It also shares similarities with other numerical linear algebra methods \cite{Noble:2013gen,Noble:2017dia}
and finds applications in analyzing PT-symmetric Hamiltonians \cite{Weideman:2006sdm,Krejcirik:2014kaa} 
and Schr\"odinger-like equations in gravitational waves \cite{Jaramillo:2020tuu,Boyanov:2022ark,Sarkar:2023rhp}. 
The pseudospectrum is a set of values at which the norm of the operator resolvent is large, 
providing insights into the stability and behavior of the system beyond its eigenvalues alone. 
The core concept behind SCM for estimating the spectrum of a differential operator is 
to replace the differential operator with a matrix. 
This replacement transforms the problem of determining the spectrum of the differential operator 
into the more straightforward task of calculating the eigenvalues of the matrix. 
When Chebyshev polynomials are used as the basis functions, the resulting matrices 
are known as {\em Chebyshev differential matrices} (CDM) \cite[\S 30]{Trefethen:2020sp}.

However, the CDM is not Hermitian, which consequently disrupts the completeness of eigenstates. 
This phenomenon is associated with the presence of exceptional points \cite{Mostafazadeh:2009zz}, 
where two or more eigenvalues and their corresponding eigenvectors coalesce. 
In other words, although the original differential operator is Hermitian, 
the corresponding matrix in the spectral collocation method is non-Hermitian. 
This naturally raises two questions: 1) If the corresponding matrix is not Hermitian, 
why are the eigenvalues real? 2) How do we reconstruct the completeness of eigenstates 
in the spectral collocation method? In this work, we will address these two questions.

The paper is structured as follows: In Sec.\  \ref{sec:harm-osci}, 
we begin with the complex harmonic oscillator to extend the conventional index theory 
in 2D dynamic systems to determine the existence of closed orbits. 
We then demonstrate the non-Hermiticity of CDM and show that it possesses PT-symmetry, 
which explains why the complex harmonic oscillator has a real spectrum 
despite the Hamiltonian matrix constructed by CDM being non-Hermitian. 
Following this, we compute the metric operator and reconstruct the completeness of eigenstates in the SCM.

In Sec.\ \ref{sec:liouville}, we apply the approaches developed in Sec.\  \ref{sec:harm-osci} to the complex Liouville theory. 
Additionally, we use perturbation theory to construct the metric operator for the Liouville model. 
This operator is essentially a shift that moves the coordinate into the imaginary infinity 
and is effective for extended models of Liouville theory, as discussed in Sec\ \ref{sec:ext} . 
In Sec\ \ref{sec:ext}, we also show how the instability of the spectrum can be affected by the compactification parameter, 
and compare the spectra of polynomial potentials and potentials with additional Liouville terms. 
Sec.\ \ref{sec:conclusion} contains the conclusions and outlook, 
followed by an appendix that provides a brief introduction to the perturbative method for calculating the metric operator.

\section{Warm-up with complex harmonic oscillator}
\label{sec:harm-osci}

In this section, we revisit the complex harmonic oscillator by 
extending the conventional index theory from phase portraits in real dynamical systems 
to the complex domain. Here, the phase portrait transforms into two sheets of Riemann surfaces. 
Our primary objective is to ascertain the presence of closed orbits, 
which is crucial for determining the existence of discrete spectra. 
Additionally, we emphasize the pseudo-hermiticity of the differential matrix within the spectral method. 
This approach allows us to reproduce the pseudospectrum and demonstrate the completeness 
of the eigenvectors associated with the complex harmonic oscillator.

\subsection{Index theory on the Riemann surfaces}
\label{sec:index}

In Newtonian mechanics, the dynamics of a one-dimensional  system are described by the equation, 
\begin{equation}
\label{eq:firstInt-Liouville}
\dot z(t) = 2\sqrt{E-V(z)},
\end{equation}
where the coordinate $z$ has been analytically continued into the complex plane 
according to the setting of the PT symmetry theory \cite{Bender:2023cem}. 
This equation can be deconstructed into its real and imaginary components by expressing $z$ as $x + \mi y$,
\begin{equation}
\label{eq:sys-liouville}
\dot x(t) = \Re\left[2\sqrt{E-V(z)}\right],\quad
\dot y(t) = \Im\left[2\sqrt{E-V(z)}\right].
\end{equation}
Consequently, the one-dimensional complex system transforms into a two-dimensional system.
However, the conventional methodologies used for analyzing two-dimensional real systems
 are not directly applicable due to the multivalued nature of the square root function.
 For instance, index theory faces challenges in addressing these dynamics. 
To elucidate this issue, consider a specific case where  $V = z^2$.
According to Ref.\ \cite{Bender:2014bna}, 
the presence of discrete eigenvalues in quantum systems depends on 
whether the orbits of the classical correspondence are closed. 
Therefore, we will utilize index theory to examine the closed orbits. 

Initially, the dynamic system yields two fixed points, $z = \pm E$, derived from solving $\sqrt{E-z^2}=0$.
Moreover, a theorem in conventional $2$-D dynamics states that any closed orbit within the phase portrait 
must encompass fixed points whose indices sum to $+1$ \cite[\S 6.8]{Strogatz:2024ndc}. 
Specifically, the indices for three types of fixed points are as follows: $I=+1$ for stable and unstable points, and 
$I=-1$ for a saddle point. Consequently, a closed orbit cannot enclose an even number of fixed points.
To illustrate this statement, let's assume the numbers of stable, unstable, and saddle points within the 
closed orbit are denoted as $n$, $m$, and $p$, respectively. The theorem dictates that $n+m-p=1$. 
Additionally, if there is an even number of fixed points in the closed orbits, then $n+m+p=2 N$, where $N$ is an integer.
From these two equations, we can deduce that $p=(2N-1)/2$. However, since $p$ must be an integer, 
this results in a contradiction. Therefore, it implies that there cannot be an even number of fixed points in a closed orbit.

Nevertheless, all the closed orbits in the complex harmonic oscillator must encircle both fixed points \cite{Bender:1998gh}. 
This appears to challenge the conventional predictions of the index theorem, as we've demonstrated before. 
This discrepancy arises from the altered configuration of the phase portrait when coordinates are analytically 
continued from the real to the complex domain. 
Specifically, the phase portrait transitions into a two-sheeted Riemann surface, 
with fixed points now acting as branch points due to the multivalued nature of the square root function.

\begin{figure}[!ht]
     \centering
         \includegraphics[width=.75\textwidth]{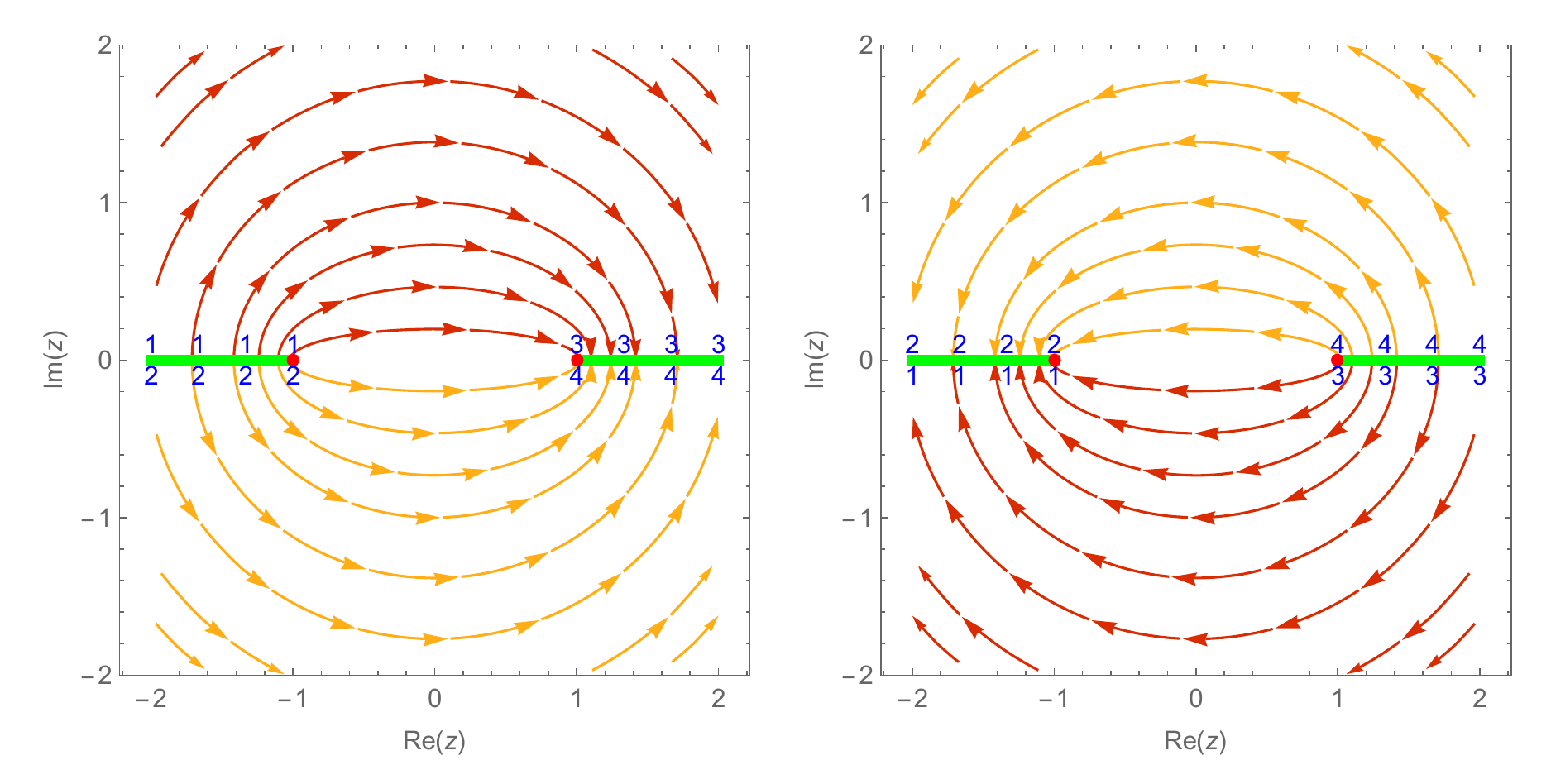}
      \captionsetup{width=.9\textwidth}
       \caption{Two-sheeted phase portrait of the complex oscillator. 
       The left panel represents the upper sheet, 
       while the right panel shows the lower sheet. 
       The bright green lines indicate the branch cuts, 
       and the two branch points are depicted by red dots. 
       The numbers indicate how the two sheets of the Riemann surface are connected, 
       with edges sharing the same numbers being glued together.}
        \label{fig:complex-oscillator}
\end{figure}

Fig.\ \ref{fig:complex-oscillator} illustrates the two-sheeted phase portrait of the complex oscillator. 
The left one represents single-valued sheet for $\sqrt{E-z^2}$, 
whereas the right one depicts $-\sqrt{E-z^2}$. 
We choose branch cuts such that they reside within the set
$\{z | \Im\left[ z^2\right]=0\}$.
The streams of identical color depict two distinct types of closed orbits. 
Each type extends across both sheets of the Riemann surface, 
and no intersection occurs between the closed orbits of different types.

\begin{figure}[!ht]
     \centering
     \begin{subfigure}[b]{0.4\textwidth}
         \centering
         \includegraphics[width=\textwidth]{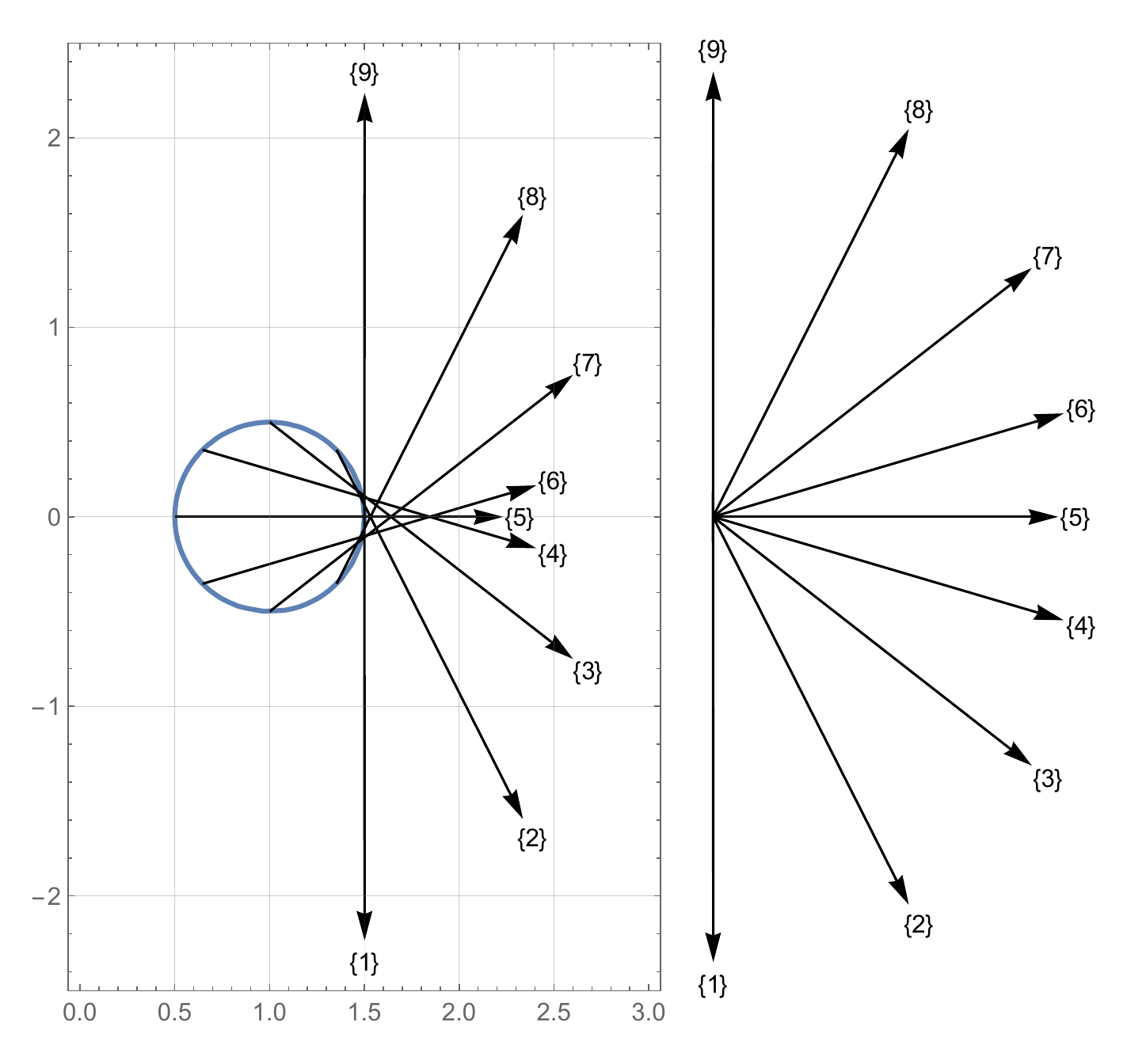}
         \caption{Upper sheet.}
         \label{fig:upper-index}
     \end{subfigure}
     \begin{subfigure}[b]{0.4\textwidth}
         \centering
         \includegraphics[width=\textwidth]{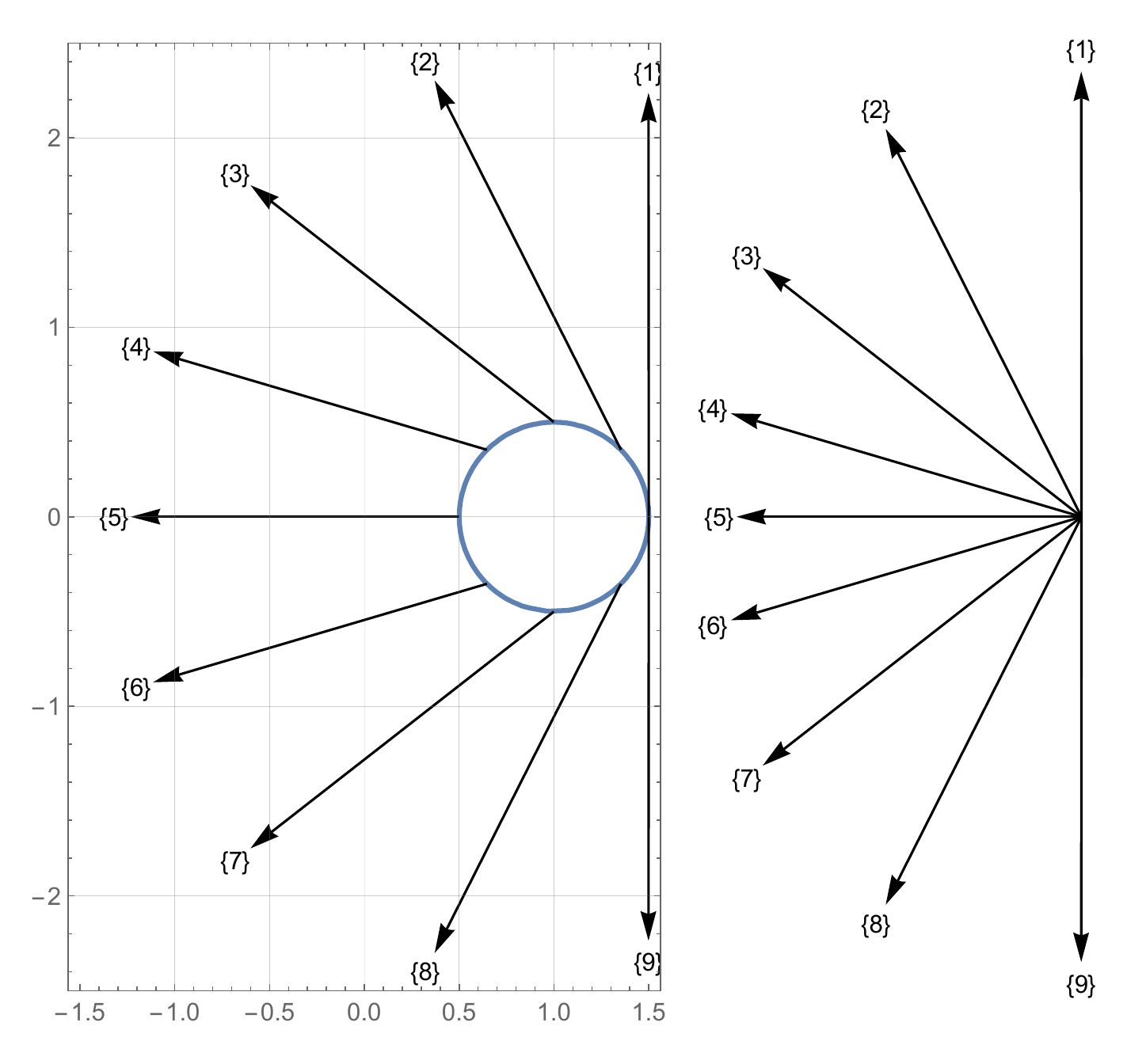}
         \caption{Lower sheet.}
         \label{fig:lower-index}
     \end{subfigure}
      \captionsetup{width=.9\textwidth}
       \caption{Schematic illustration for estimating the index of the fixed point on the right. The detailed methodology is described in Ref.\ \cite[\S 6.8]{Strogatz:2024ndc}.}
        \label{fig:index-schem}
\end{figure}

We now proceed to analyze the indices of the system. 
Focusing on the fixed point on the right, we begin 
by plotting two nearly closed circles, $C$, on each of the two sheets of the Riemann surface, centering 
them on this fixed point. 
Due to the presence of the branch cut, each circle on the respective sheet will 
feature a slight gap at the location of the cut. 
It is crucial to select a radius that is not so large as to 
intersect the cut line extending from the left fixed point. 
Next, we trace a selection of vectors along the 
corresponding streamlines on these circles, as depicted in the schematic illustration provided in Fig.\ \ref{fig:index-schem}.

In the upper sheet of the Riemann surface (refer to Fig.\ \ref{fig:upper-index}), 
the vector, initiating from the position labeled $\{1\}$, traverses the circle in a counterclockwise direction, 
transitioning to the position labeled $\{9\}$ after accumulating a total rotation of $\pi$. 
Consequently, for the upper sheet of the Riemann surface, the index of the right fixed point is given by
\begin{equation}
I_C= \frac{1}{2\pi} [\phi]_C = \frac{1}{2}
\end{equation}
Similarly, in the lower sheet of the Riemann surface (refer to Fig.\ \ref{fig:lower-index}), the vector 
completes a circuit around the circle, evolving from $\{1\}$ to $\{9\}$, with a total rotation of $\pi$.
Thus, in the lower sheet, the index of the right fixed point is also a fractional value, $1/2$, 
which differs from the index theory in traditional dynamics systems \cite{Strogatz:2024ndc}.

Although the indices of the fixed points have transitioned from conventional integers to fractions, 
the index theorem (Theorem 6.8.2 in \cite{Strogatz:2024ndc}) remains applicable in this context. 
For the complex harmonic oscillator, the closed orbits must encircle both fixed points for the sum of their 
indices to equal one. As previously mentioned, the closed orbits in this scenario traverse a 
two-sheeted Riemann surface and are divided into two distinct classes that do not intersect, as depicted 
in Fig.\ \ref{fig:complex-oscillator}.

Furthermore, the introduction of fractional indices is accompanied by a change in the nature of fixed 
points across different sheets of the Riemann surface. 
For instance, the point where $x=E$ on the upper sheet is ``stable'';
however, it transforms into ``unstable'' on the lower sheet. 
Notably, the branch cut originating from the point $x=E$ exhibits distinct attributes on both the upper and lower sheets of 
the Riemann surface. On the upper sheet, all trajectories converge towards the branch cut, endowing 
it with stability. Consequently, the fixed point at the termination of the branch cut is inherently stable, 
although it is important to highlight that only a single trajectory converges to this fixed point. 
We term such fixed points with these characteristics as {\em quasi-stable} because they deviate from the conventional criteria of stability.

We conclude this section by commenting on the relationship between discrete spectra and the existence of closed orbits. 
On the Riemann surfaces of the harmonic oscillator, 
orbits can be classified into two types based on their trajectories: 
closed orbits spanning both sheets of the Riemann surface, 
as discussed earlier, involving both counterclockwise and clockwise rotations around the two fixed points simultaneously, 
and oscillatory motions between the two fixed points. The former type of orbit leads to discrete real energy spectra  \cite{Bender:1998gh}, 
while the latter undoubtedly contributes to energy spectra with similar characteristics. 
This suggests that besides closed orbits, the presence of a trajectory linking any two fixed (branch) points 
must also be considered as a criterion for the existence of a discrete real energy spectrum.

\subsection{Pseudo-hermiticity of the differential matrices}

The pseudospectrum of an operator $A$ consists of complex numbers $z$ for which $A-z I$ has a large resolvent $\| (A- z I)^{-1}\|$. 
The term ``large''  is determined by a parameter $\epsilon$. 
Specifically,  for any given $\epsilon$, the $\epsilon$-pseudospectrum of an operator $A$, 
denoted by $\sigma_\epsilon(A)$, is the set of $z\in\mathbb{C}$ such that $\| (A- z I)^{-1}\|>\epsilon^{-1}$.
SCM related to the pseudospectrum has been introduced in physics to estimate the spectra \cite{Weideman:2006sdm,Krejcirik:2014kaa} 
and investigate the spectral instability of non-normal operators \cite{Jaramillo:2020tuu,Boyanov:2022ark,Destounis:2023nmb}.

The essence of this method lies in converting the Hamiltonian with differential operators into matrices. 
In this paper, we utilize the Chebyshev functions as bases, transforming the differential operators of spatial coordinates 
into the CDM.
For an integer $N\ge 1$, the CDM $D_N$ has dimension $(N+1) \times (N+1)$. 
It can be depicted by a block matrix
\begin{equation}
D_N=
\begin{bmatrix}
(D_N)_{00} &   (D_N)_{0j}   & (D_N)_{0N} \\
(D_N)_{i0}  &   (D_N)_{ij}   &   (D_N)_{iN} \\  
(D_N)_{N0} &   (D_N)_{Nj}   &  (D_N)_{NN} \\
\end{bmatrix},
\end{equation}
where all elements are {\em real} numbers
\begin{subequations}
\begin{equation}
(D_N)_{00} = \frac{2N^2+1}{6},\qquad  (D_N)_{NN} = -\frac{2N^2+1}{6},
\end{equation}
\begin{equation}
(D_N)_{jj} = \frac{-x_j}{2(1-x_j^2)},\qquad  j = 1,\ldots, N-1,
\end{equation}
\begin{equation}
(D_N)_{ij} = \frac{c_i}{c_j} \frac{(-1)^{i+j}}{(x_i-x_j)},\quad
i\neq j,\quad i, j = 0,\ldots, N,
\end{equation}
\end{subequations}
The $x_j$ represent Chebyshev interpolation points
\begin{equation}
x_j = \cos(j\pi /N),\quad j=0,1,\ldots, N,
\end{equation}
and the constants $c_i$ are given as follows
\begin{equation}
c_i =
\begin{cases}
2 & i = 0 \text{ or } N,\\
1 & \text{otherwise}.
\end{cases}
\end{equation}

Thus, the Hamiltonian operator of the harmonic oscillator takes the form
\begin{equation}
H_{\rm Cheb}=-\widetilde{D}_N^{2} +V_0 \diag \left\{
x_1^2,\ldots
 x_j^2,\ldots
x_{N-1}^2
\right\}.
\end{equation}
Here $\widetilde{D}_N^{2}$ denotes the second-order CDM,
which has dimensions of $(N-1)\times(N-1)$.
This matrix is derived by removing the first and last rows and columns of
$D_N^{2}$ to account for the boundary conditions \cite[\S 7]{Trefethen:2000smt}.

However, $H_{\rm Cheb}$ is not symmetric (Hermitian) because $(\widetilde{D}_N^{2})_{ij}$ is not symmetric matrix.
To demonstrate this aspect of the CDM, we explicitly express $(\widetilde{D}_N^{2})$ as
\begin{equation}
(\widetilde{D}_N^{2})_{ij}=\sum_{k=0}^N(D^N)_{ik}(D^N)_{kj},\quad
i, j = 1,\ldots, N-1.
\end{equation}
The matrix $\widetilde{D}_N^{2}$ can be decomposed into three components
\begin{equation}
\widetilde{D}_N^{2}=C^2
+A
+B,
\end{equation}
where each component is defined as follows
\begin{subequations}
\begin{equation}
C_{ij} = (D^N)_{ij},\quad
i, j = 1,\ldots, N-1,
\end{equation}
\begin{equation}
A_{ij} = (D^N)_{i0}(D^N)_{0j}=-\frac{(-1)^{i+j}}{(1-x_i)(1-x_j)},
\end{equation}
and
\begin{equation}
B_{ij} = (D^N)_{iN}(D^N)_{Nj}=-\frac{(-1)^{i+j}}{(1+x_j)(1+x_i)}.
\end{equation}
\end{subequations}

Furthermore,  $C$ can be decomposed into two components
\begin{equation}
C = C_{\rm even}+ C_{\rm odd},
\end{equation}
where $C_{\rm even}$ is a diagonal matrix, 
\begin{equation}
(C_{\rm even})_{jj} =(D_N)_{jj},\quad  j = 1,\ldots, N-1,
\end{equation}
and $C_{\rm odd}$ is a hollow matrix (with all diagonal entries being zero) 
\begin{equation}
(C_{\rm odd})_{ij} =(D_N)_{ij},\quad i\neq j,\quad j = 1,\ldots, N-1.
\end{equation}
Given that $(D_N)_{ij}=-(D_N)_{ji}$ for $i\neq j$, 
the even part $C_{\rm even}$ is symmetric, and the odd part $C_{\rm odd}$ is anti-symmetric, 
\begin{equation}
\left(C_{\rm even}\right)^\intercal = C_{\rm even},\quad
\left(C_{\rm odd}\right)^\intercal = -C_{\rm odd}.
\end{equation}
Therefore, $C^2$ is not symmetric, $\left(C^2\right)^\intercal \neq C^2$, and consequently, $(\widetilde{D}_N^{2})$  is not symmetric $(\widetilde{D}_N^{2})^\intercal \neq \widetilde{D}_N^{2}$.
This implies the completeness of the eigenstate may be violated even for a Hermitian differential operator.

Although $\widetilde{D}_N^{2}$ is not symmetric matrix, 
it possesses two additional symmetries. 
One is parity symmetry: $\widetilde{D}_N^{2}(x)=\widetilde{D}_N^{2}( - x)$, 
and the other is $(\widetilde{D}_N^{2})_{ij}=(\widetilde{D}_N^{2})_{(N-1-i) (N-1-j)}$, 
where $i, j = 1,\ldots, N-1$.
The last symmetry is straightforward to verify from the definition of the CDM.
To observe the parity symmetry, note that $A$ and $B$ are transformed into each other under the parity transformation $x\to - x$, 
implying that the sum $A+B$ has parity symmetry. 
Meanwhile, the $C^2$ can be separated into four parts
\begin{equation}
C^2 = C_{\rm even}^2 +C_{\rm odd}^2+C_{\rm even} C_{\rm odd} +C_{\rm odd} C_{\rm even},
\end{equation}
where 
\begin{subequations}
\begin{equation}
(C_{\rm even}^2 )_{jj}= \frac{x_j^2}{4(1-x_j^2)^2},\quad
(C_{\rm odd}^2 )_{ij}=  \sum_{k}\frac{(-1)^{i+j}}{(x_i-x_k)(x_k-x_j)}, 
\end{equation}
\begin{equation}
(C_{\rm even} C_{\rm odd})_{ij}=\frac{-x_i}{2(1-x_i^2)}\frac{(-1)^{i+j}}{(x_i-x_j)},
\end{equation}
\begin{equation}
(C_{\rm odd} C_{\rm even})_{ij}= \frac{(-1)^{i+j}}{(x_i-x_j)} \frac{-x_j}{2(1-x_j^2)}.
\end{equation}
\end{subequations}
Since each part is symmetric under the parity transformation $x\to - x$, $C^2$ exhibits parity symmetry.
Considering the entries of $\widetilde{D}_N^{2}$ are real, 
$\widetilde{D}_N^{2}$ exibits both time-reversal and parity symmetries. 
Hence, the eigenvalues of $\widetilde{D}_N^{2}$ are real according to the principles of PT-symmetric quantum mechanics \cite{Bender:2023cem}.

\begin{figure}[!ht]
     \centering
     \begin{subfigure}[b]{0.48\textwidth}
         \centering
         \includegraphics[width=\textwidth]{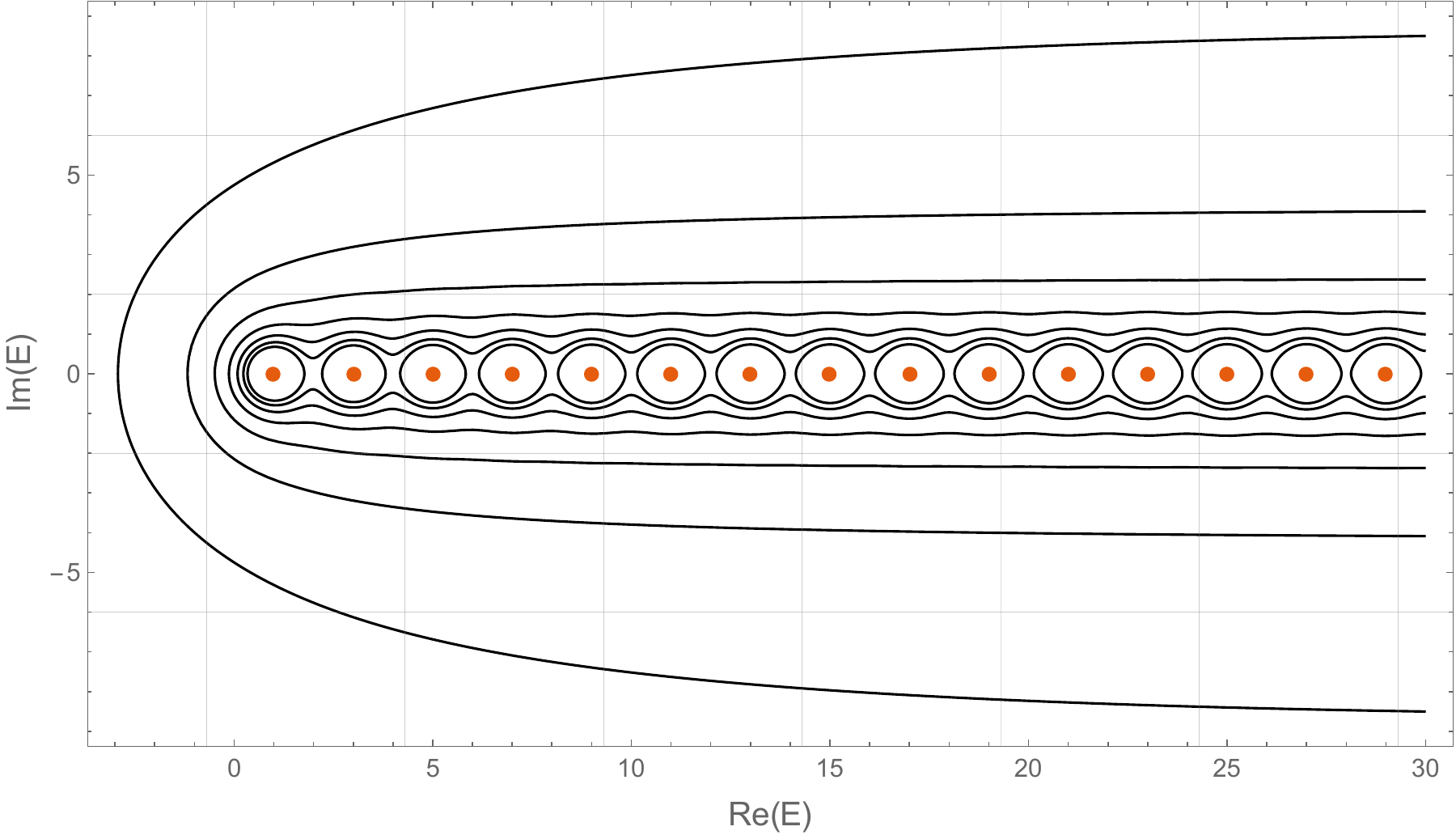}
         \caption{Pseudospetrum.}
         \label{fig:oscillator-en}
     \end{subfigure}
     \begin{subfigure}[b]{0.42\textwidth}
         \centering
         \includegraphics[width=\textwidth]{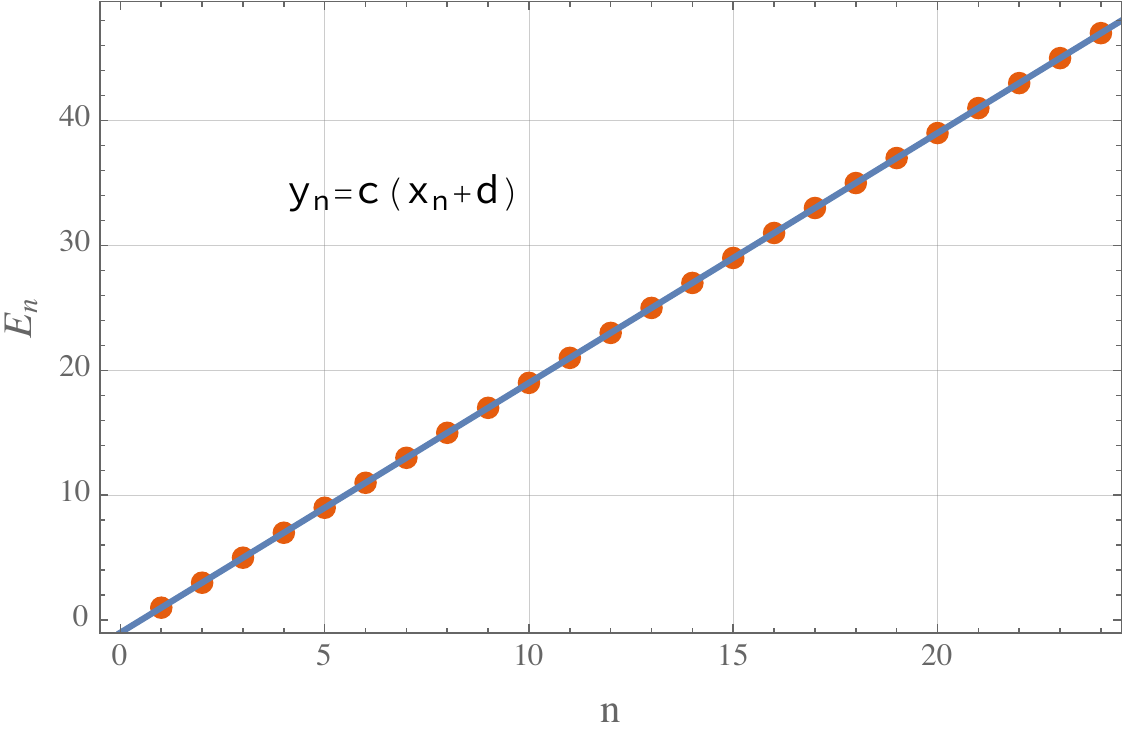}
         \caption{Fitting of discrete energy.}
         \label{fig:oscillator-fit}
     \end{subfigure}
      \captionsetup{width=.9\textwidth}
       \caption{The pseudospectrum of harmonic oscillator. The left one shows the pseudospectrum (the black counters) and
       the spectrum (orange points); the right one provides the linear fitting of the spectrum with respect to $n$.}
        \label{fig:oscillator}
\end{figure}

Fig.\ \ref{fig:oscillator-en} illustrates the pseudospectrum of the harmonic oscillator with  $N=70$ 
and a symmetric compactification $x\in[-8, 8]$. In the plot, the orange points represent the eigenvalues in the complex plane of $E$, 
while the contours depict the $\epsilon$-pseudospectrum \cite{Trefethen:2020sp,Trefethen:2000smt}, denoted as $\Lambda_\epsilon(H_{\rm Cheb})$,
 with seven arbitrary values of $\epsilon$. This pseudospectrum is defined as
\begin{equation}
\Lambda_\epsilon(H_{\rm Cheb}) 
=\left\{
z\in\mathbb{C}\big |
\left \|(z I - H_{\rm Cheb})^{-1}\right\|\ge \epsilon^{-1}
\right\},
\end{equation}
where $\|\cdot\|$ represents the matrix 2-norm, and $I$ is identity matrix with the same dimension as $H_{\rm Cheb}$.
Fig.\ \ref{fig:oscillator-fit}  displays the fitting of spectral points. 
The fitting equation, $E_n = c ( n + d)$, where $n$ starts from $1$, accurately describes the numeric data. 
Notably, the fitting parameters, $c=2.00002$ and $d=-0.500067$, 
closely match those commonly found in standard textbooks. 
This alignment validates the consistency and reliability of the analysis.

To construct the Hermitian version of $H_{\rm Cheb}$ and ensure the completeness of eigenstates, 
we require the metric operator \cite{Mostafazadeh:2002pht}, which can be calculated using the generalized method \cite{Feinberg:2021npz}, see also App.\ \ref{app:pert}. 
However, since all eigenvalues of the CDM are real and positive 
(indicating that the CDM is positive-definite), the process is simplified. 
We have
\begin{equation}
\begin{split}
H_{\rm Cheb}^{\dagger } & =\left(S^{-1}\Lambda  S\right)^{\dagger }
= S^{\dagger }\Lambda  (S^{-1})^{\dagger }\\
& = S^{\dagger }S H_{\rm Cheb}S^{-1}(S^{-1})^{\dagger }
=S^{\dagger }S H_{\rm Cheb}(S^{\dagger }S)^{-1},
\end{split}
\end{equation}
where $\Lambda$ represents the spectrum of $H_{\rm Cheb}$ and is a diagonal matrix. 
Meanwhile, $S$ is constructed from the column eigenvector, $S=[v_1,\ldots, v_{N-1}]^{-1}$. 
Thus, the metric operator is given by
\begin{equation}
\eta=S^{\dagger }S.
\end{equation}

\begin{figure}[!ht]
     \centering
         \includegraphics[width=.6\textwidth]{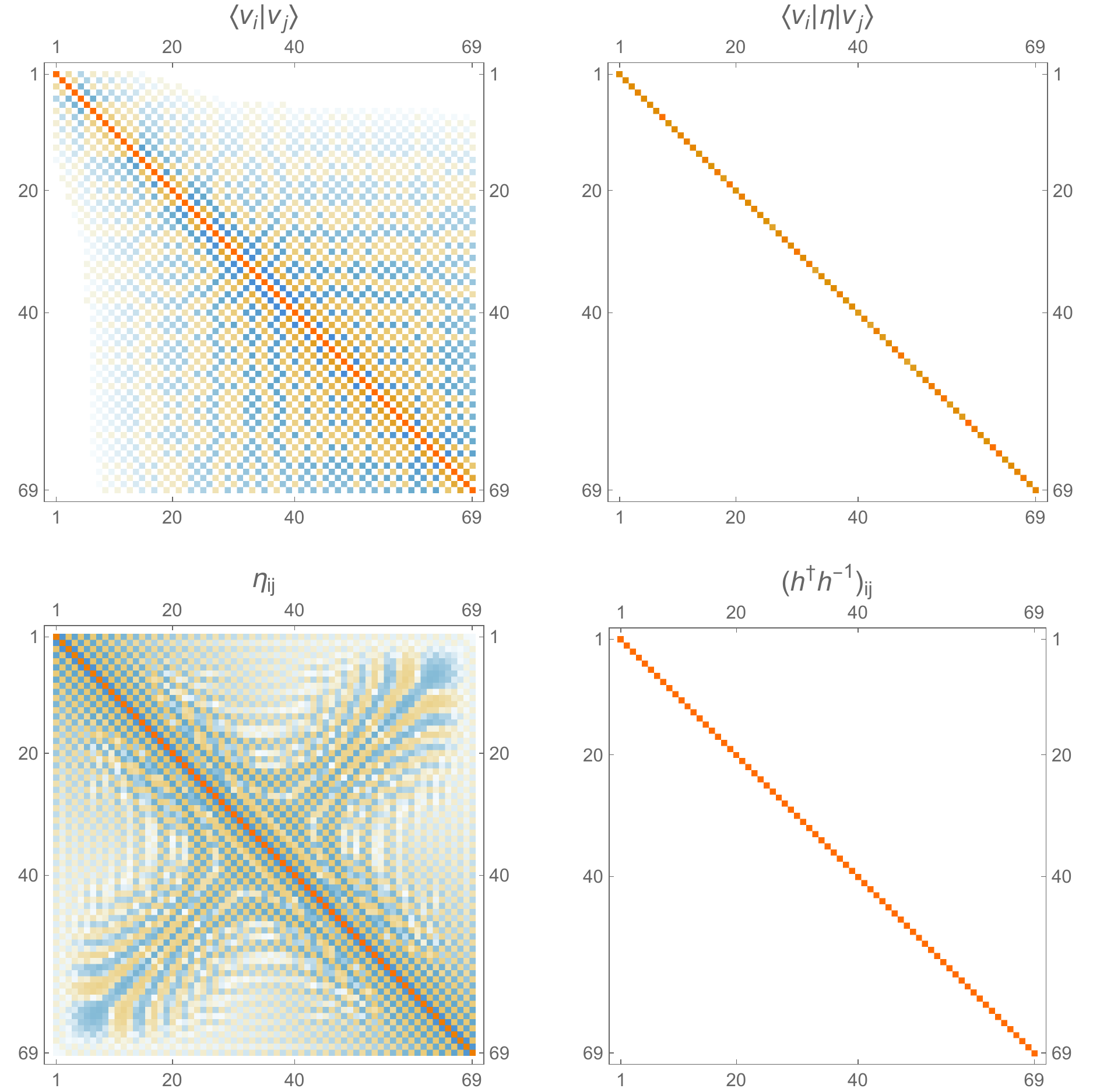}
      \captionsetup{width=.9\textwidth}
       \caption{Comparison of eigenvector completeness and Hermitian Hamiltonian.}
        \label{fig:oscillator-complete}
\end{figure}

To demonstrate the completeness of the eigenvectors, 
we plot the inner product as a matrix with an accuracy of $0.1^{69}$ in Fig.\ \ref{fig:oscillator-complete}.
In this representation, each matrix element is represented by a unique color. 
The diagonal elements, highlighted in orange-red, correspond to a matrix value of unity, while the blank regions indicate matrix elements that are zero, 
thus being able to visualize the completeness of the eigenvectors.
The plot in the first row, left panel, shows the ill-defined completeness,
while the second panel displays the 
corrected completeness achieved by introducing the metric operator $\eta$.
In the second row, the left panel depicts the metric operator $\eta_{ij}$, 
and the right panel illustrates the Hermiticity of the Hamiltonian $h$.
The  Hermitian equivalence of the Hamiltonian can be constructed as $h = \rho H_{\rm Cheb} \rho^{-1}$, 
where $\rho = \sqrt{\eta}$. This Hermitian Hamiltonian $h$ should provide the same spectrum as $H_{\rm Cheb}$.

In summary, the SCM using the non-Hermitian $H_{\rm Cheb}$ matrix yields a spectrum that closely matches the one directly obtained from the Hermitian differential operator. However, the eigenstates produced by this method fail to maintain completeness, which precludes the description of observables in terms of these states. 
Moreover, SCM not only offers an efficient method of calculating eigenvalues but also facilitates the reconstruction of completeness, making it a more versatile tool in this context.

\section{PT-symmetric Liouville quantum mechanics}
\label{sec:liouville}

Let us now turn to the PT-symmetric Liouville quantum mechanics, whose classical Hamiltonian is given by 
\begin{equation}
\label{eq:liouville}
H = p^2 +V_0 \me^{2 \mi z},
\end{equation}
where $V_0$ is a real parameter. 
We will first examine the classical dynamics, followed by an exploration of the quantum theory. 
At the quantum level, we will determine its Hermitian equivalence using the perturbation theory 
and analyze the energy spectrum and stability using the pseudospectrum approach.

\subsection{Classical dynamics}

\begin{figure}[!ht]
     \centering
     \begin{subfigure}[b]{0.45\textwidth}
         \centering
         \includegraphics[width=\textwidth]{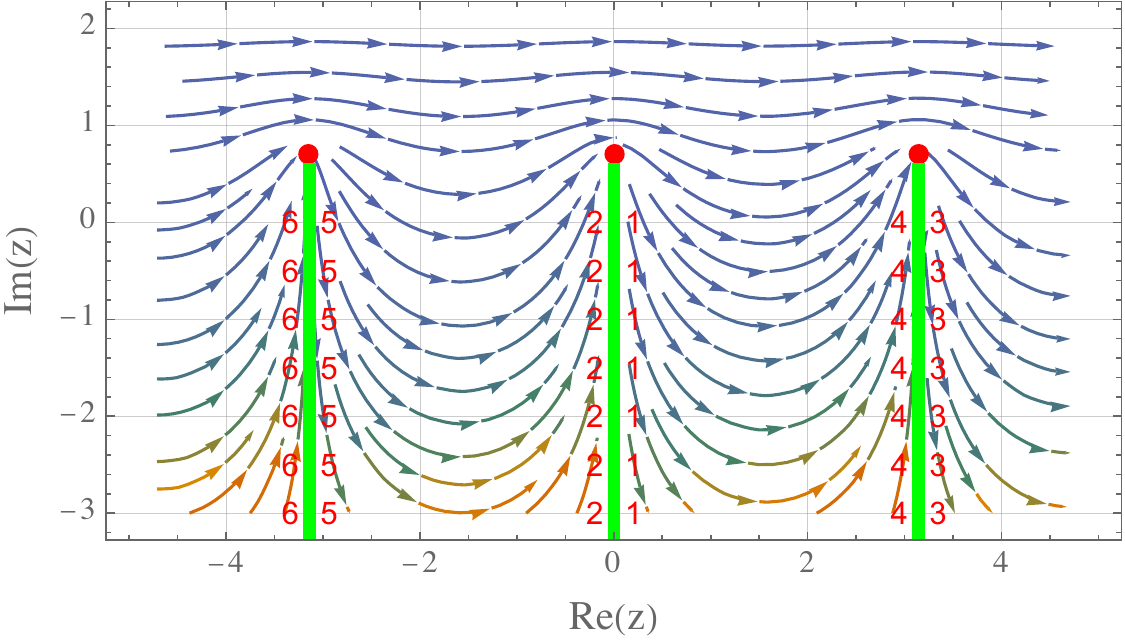}
         \caption{Upper sheet.}
         \label{fig:class-exp2ix-1}
     \end{subfigure}
     \begin{subfigure}[b]{0.45\textwidth}
         \centering
         \includegraphics[width=\textwidth]{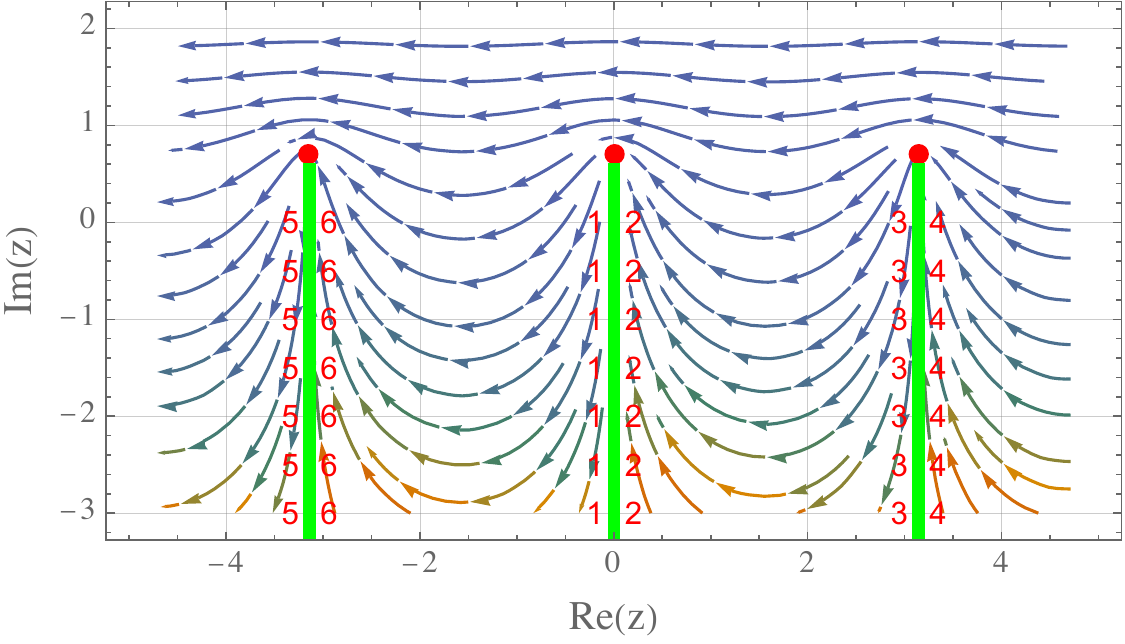}
         \caption{Lower sheet.}
         \label{fig:class-exp2ix-2}
     \end{subfigure}
      \captionsetup{width=.9\textwidth}
       \caption{
       Phase portrait of Eq.\ \eqref{eq:sys-liouville} on a two-sheeted Riemann surface. 
       The red points indicate the system's fixed points. 
       The bright green lines represent the branch cuts, 
       while the numbers show how the two sheets of the Riemann surface are connected; 
       edges with matching numbers are joined together.}
        \label{fig:class-exp2ix}
\end{figure}

From the first integral of motion, we derive the dynamic equation
\begin{equation}
\label{eq:firstInt-Liouville}
\dot z(t) = 2\sqrt{E-V_0 \me^{2\mi z(t)}}.
\end{equation}
where the fixed points, found by solving $\sqrt{E-V_0 \me^{2\mi z_c}}=0$, 
form an infinite number due to the periodicity of the Liouville potential
\begin{equation}
z_c= k \pi -\frac{\mi}{2}  \ln \left(E/V_0\right),\quad k\in \mathbb{Z},
\end{equation}
with each fixed point having a fractional index $1/2$.
To further analyze the system using the method we developed above, 
we transform the dynamic equation into a $2$D  system
by substituting $z(t)=x(t)+\mi y(t)$ into the Eq.\ \eqref{eq:firstInt-Liouville}, 
yielding
\begin{equation}
\label{eq:sys-liouville}
\dot x(t) = \Re\left[2\sqrt{E-V_0 \me^{2\mi z(t)}}\right],\quad
\dot y(t) = \Im\left[2\sqrt{E-V_0 \me^{2\mi z(t)}}\right].
\end{equation}
The phase portrait is shown in Fig.\ \ref{fig:class-exp2ix}, 
where the two sheeted Riemann surfaces arise due to the multivalued nature of the square root function.
The branch cuts are chosen such that they reside within the set
$\{z | \Im\left[V(z)\right]=0\}$.

Firstly, unlike the complex harmonic oscillator, each orbit in this system remains on only one sheet, 
despite the presence of two sheets in the Riemann surface. 
That means that if the initial position is on the upper (or lower) sheet, 
the ``classical particle'' will continue to move exclusively on the upper (or lower) sheet and will not cross the branch cut to the other sheet. 
Thus, the orbits on these two sheets of the Riemann surface are entirely independent of each other.

Secondly, there are no closed orbits in this system. 
Any orbit that would encircle two fixed points (see the index theory in Sec.\ \ref{sec:index}) 
would need to cross the branch cuts and transition to the other sheet of Riemann surface, 
which does not occur. Furthermore, unlike the complex harmonic oscillator, 
there are no bounded trajectories that terminate at two fixed points. 
Instead, any trajectory starting at a fixed point will ultimately tend towards negative imaginary infinity.
Therefore, to acquire a discrete energy spectrum, it is necessary to implement a compactification operation. 
Given the specific nature of the Liouville model, we can incorporate a periodic compactification parameter. 
This parameter serves to confine the potential energy, ensuring the discreteness of the spectrum.
But before that, let us investigate the Hermitian equivalence of the model.


\subsection{Equivalent Hermitian Hamiltonian}


To derive the equivalent Hermitian Hamiltonian in PT-symmetric Liouville quantum mechanics, 
we use the perturbation technique outlined in App.\ \ref{app:pert}. 
We begin by introducing an infinitesimal parameter 
$\e$  and decompose the Hamiltonian from Eq.\ \eqref{eq:liouville} as follows
\begin{equation}
\label{eq:modliouville}
H^\e= H_0+\e H_1,
\end{equation}
where $H_0= p^2 +V \cos(2 z)$ is Hermitian and
$H_1=\mi  V \sin(2 z)$ is anti-Hermitian. 
Next, we employ Eq. \eqref{eq:pert-Q} to find particular solutions for  $Q_{2n-1}$,  yielding
\begin{equation}
Q_{2n-1}=-\frac{p}{2n-1},\quad
n=1,2,3,\ldots.
\end{equation}
By summing the perturbative series, 
we construct the operator $Q^\e$
\begin{equation}
\label{eq:shift}
Q^\e = \sum_{n=1}^\infty Q_{2n-1} \e^{2n-1}
=-p\sum_{n=1}^\infty \frac{ \e^{2n-1}}{2n-1}
=-p \arctanh(\e),
\end{equation}
The coefficients of the perturbatively equivalent Hamiltonian can then be expressed in a closed form as
\begin{equation}
h_0 =p^2 + V \cos(2z),\quad
h_{2n} =-\frac{\Gamma (n-1/2)}{2\sqrt{\pi}\Gamma(n+1)}V \cos(2z),
\quad n=1, 2, 3,\ldots.
\end{equation}
This results in the following sum
\begin{equation}\begin{split}
h^\e &=\sum_{n=0}^\infty h_{2n} \e^{2n}
=p^2 + V \cos(2z)
-V \cos(2x)\sum_{n=1}^\infty \frac{\Gamma (n-1/2)}{2\sqrt{\pi}\Gamma(n+1)}\e^{2n}\\
&=p^2+V\sqrt{1-\e^2} \cos(2z).
\end{split}\end{equation}
Finally, as $\e$ approaches unity, we obtain the limit
\begin{equation}
\label{eq:eq-her}
h=\lim_{\e\to 1} h^\e 
= \lim_{\e\to 1} \left[p^2+V\sqrt{1-\e^2} \cos(2z)\right]= p^2,
\end{equation}
which shows that the Hamiltonian remains well-behaved, while $Q^\e$ becomes singular as $\e\to 1$, i.e.
\begin{equation}
\label{eq:infty}
Q=\lim_{\e\to 1} Q^\e=-p \lim_{\e\to 1} \arctanh(\e)\to -\infty.
\end{equation}
Thus, the operator for similarity transformation, $\rho^\e=\me^{Q^\e/2}$,  
can be interpreted as a shift operator along the imaginary axis. 
The divergence in Eq.\ \eqref{eq:infty} indicates that the shift must extend to imaginary infinity. 
This insight allows us to obtain the equivalent Hermitian Hamiltonian by substituting
$z\to z+\frac{\mi}{2} \arctanh(\e)$ into \eqref{eq:modliouville} directly and then taking the limit $\e\to1$.
Eq.\ \eqref{eq:eq-her} suggests that Liouville quantum mechanics is equivalent to the behavior of a free particle. 
However, when we introduce periodic compactification, Liouville quantum mechanics is equivalent to the dynamics of a free particle confined to a circular path.

\subsection{Pseudospectrum analysis}

\begin{figure}[!ht]
     \centering
         \includegraphics[width=.65\textwidth]{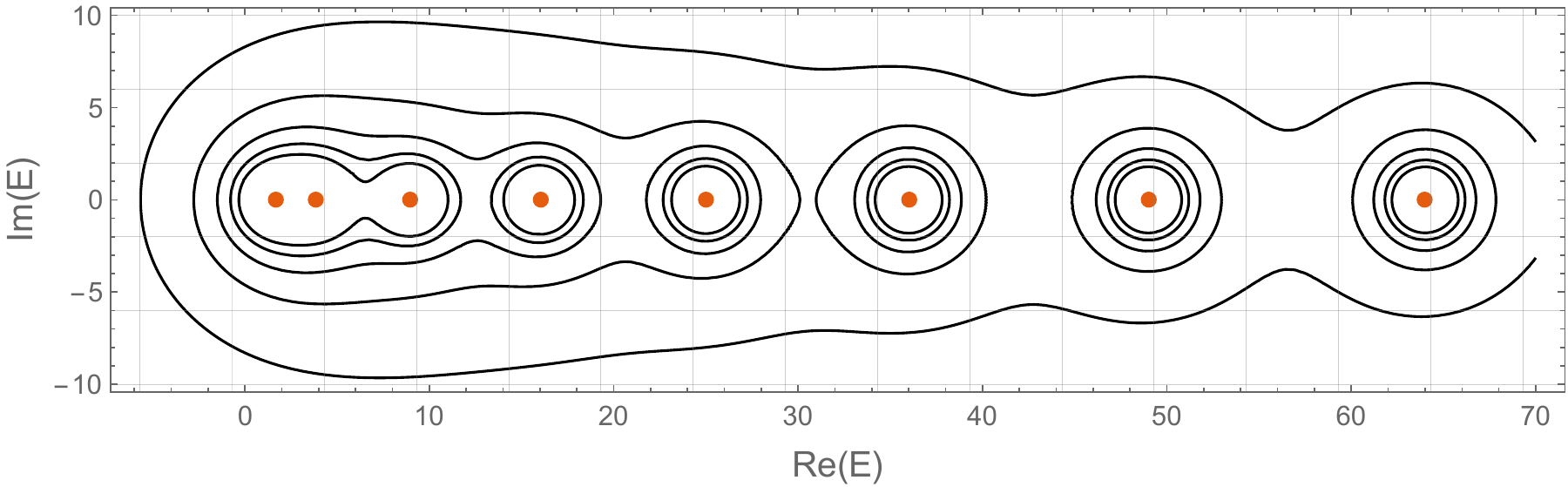}
      \captionsetup{width=.9\textwidth}
       \caption{Pseudospectrum of PT-symmetric Liouville theory, $V=V_0 \me^{2\mi x}$ with $V_0=1$ and $k=1$.}
        \label{fig:exp2ix}
\end{figure}

\begin{figure}[!ht]
     \centering
         \includegraphics[width=.56\textwidth]{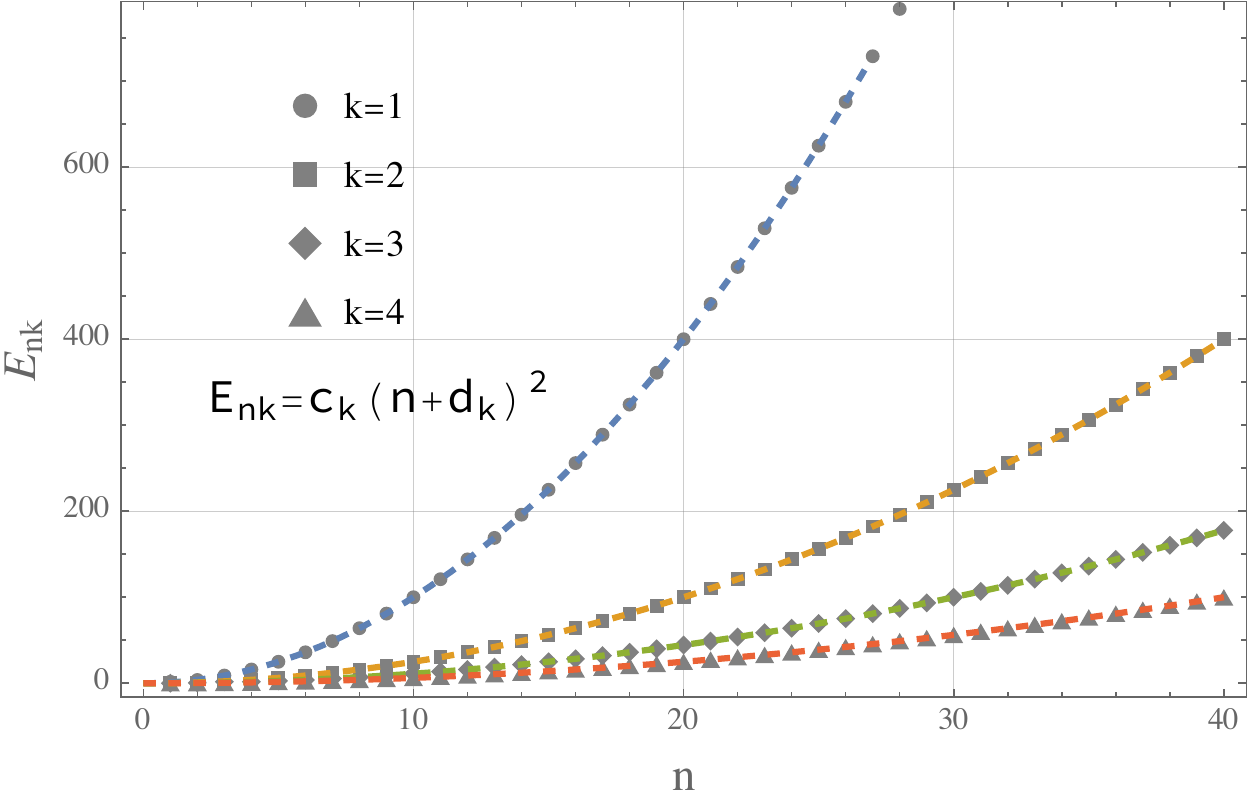}
      \captionsetup{width=.9\textwidth}
       \caption{The energy spectra for different values of $k$. The dashed curves represent fittings.}
        \label{fig:fitting-exp2ix}
\end{figure}

To derive the discrete spectrum for the Liouville quantum mechanics, 
we adopt a symmetric compactification $x\in[-L, L]$ \cite{Samsonov:2005cpn}, 
where $L=\pi k/2$ and $k\in\mathbb{Z}$.
Fig.\ \ref{fig:exp2ix} presents the pseudospectrum with  $N=70$ and $k=1$, 
The orange points indicate the eigenvalues in the complex plane of $E$, 
while the contours represent to the $\epsilon$-pseudospectrum $\Lambda_\epsilon(H_{\rm Cheb}) $ with five arbitrary values of $\epsilon$. 
In Figure \ref{fig:fitting-exp2ix}, the spectrum is represented by gray dots, and their distribution follows the trend of  $n^2$  with $n\in\mathbb{Z}$.

For more generic cases, the analytical solution for the energy spectrum is given by $E_{n k} = n^2/k^2$.
Fig.\ \ref{fig:fitting-exp2ix} also shows the numeric results for different values of $k$ in addition to the case where $k=1$.
The energy exhibits an inversely squared decreasing trend as the bounded region increases. 
As the bounded region approaches infinity, the discrete energy spectrum becomes continuous.
The fitting parameter of $c_k$ coincides with $1/k^2$, as shown in Tab.\ \ref{tab:fitting}.
\begin{table}[!ht]
\begin{center}
\begin{tabular}{c|cccc}
\hline
$k$ & 1 & 2 & 3 & 4\\
\hline
$1/\sqrt{c_k}$ & 0.99997 & 1.99985 & 3.00031 & 3.99908 \\
$d_k$ & -0.00059 & -0.00218 & 0.00330 &-0.00864 \\
\hline
\end{tabular}
\end{center}
\captionsetup{width=.9\textwidth}
\caption{Fitting parameters\label{tab:fitting}}
\end{table}%

\begin{figure}[!ht]
     \centering
         \includegraphics[width=.6\textwidth]{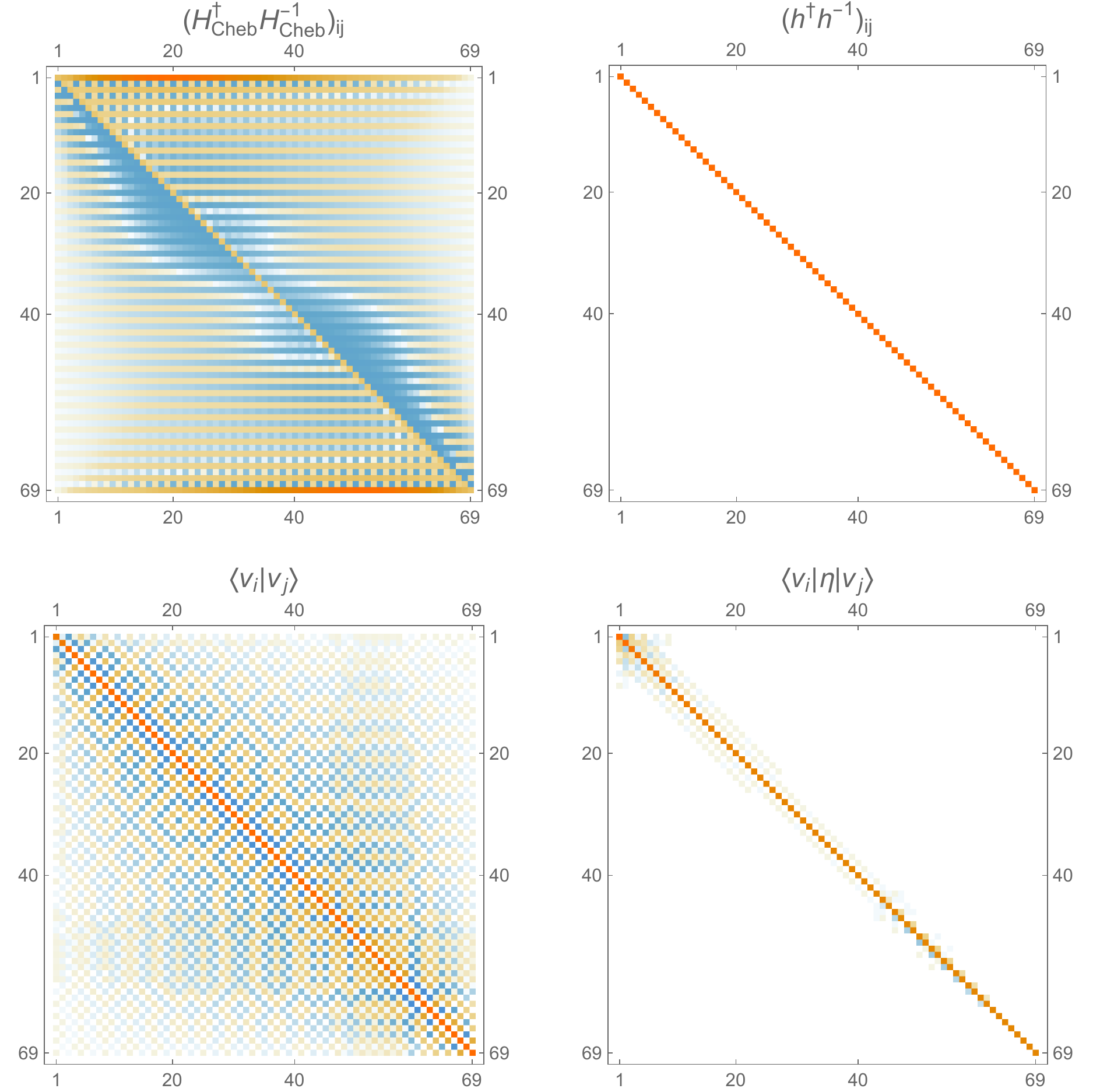}
      \captionsetup{width=.9\textwidth}
       \caption{Hermiticity and completeness of Liouville quantum mechanics. \label{fig:fexp2ix-heta}}
\end{figure}

In this model, two terms disrupt the Hermiticity in Hamiltonian: 
the CDM and the Liouville potential.
The first row of Fig.\ \ref{fig:fexp2ix-heta} shows the $H_{\rm Cheb}$'s Hermiticity breaking and its Hermitian counterpart $h$. 
The second row compares the completeness of eigenstates. 
The left plot, using the naive definition of orthogonality, clearly shows that completeness is broken. 
The right plot shows the restored completeness after introducing the metric operator.

\begin{figure}[!ht]
     \centering
         \includegraphics[width=.6\textwidth]{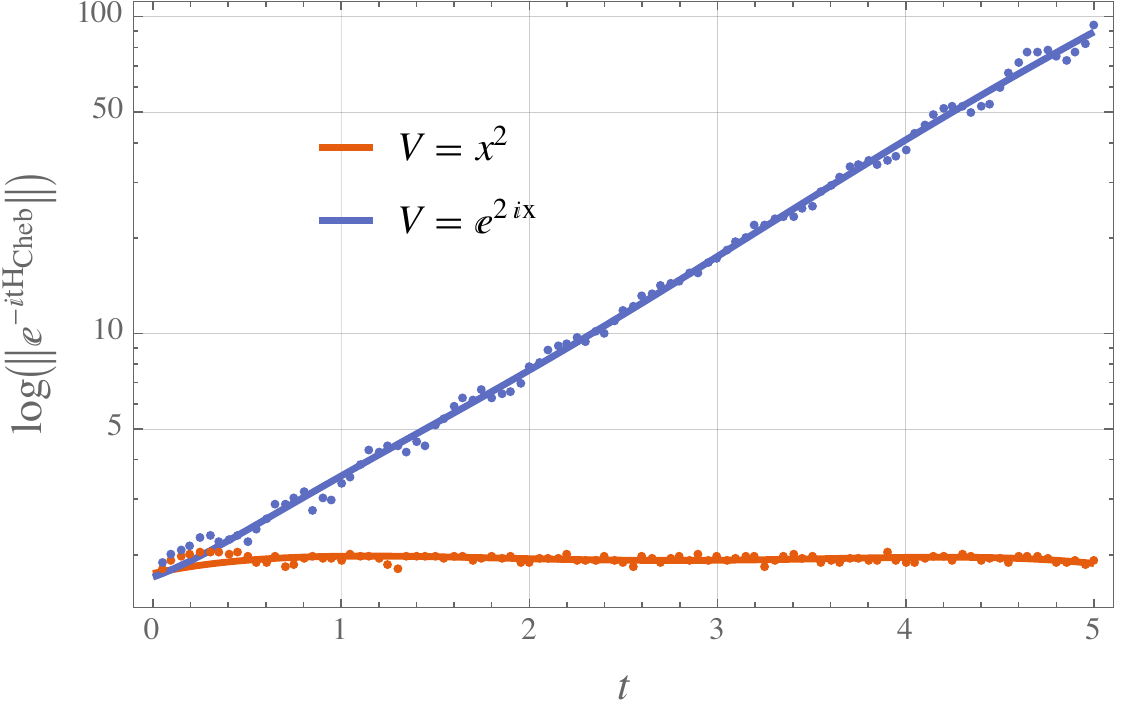}
      \captionsetup{width=.9\textwidth}
       \caption{Log-plot of evolution function $\left\| \me^{-\mi t H_{\rm Chen} }\right\|$ for Liouville model and harmonic oscillator. The dots are obtained by the numeric estimation, while the curves correspond to the fittings. }
        \label{fig:evolution}
\end{figure}

We further analyze the evolution in the traditional sense 
\begin{equation}
\frac{\dif}{\dif t}\psi = -\mi H_{\rm Chen} \psi,
\end{equation}
and examine the norm of evolution $\left\| \me^{-\mi t H_{\rm Chen} }\right\|$.
 Fig.\ \ref{fig:evolution} compares the Liouville model with the harmonic oscillator.
The harmonic oscillator shows the unitary evolution, with $\left\| \me^{-\mi t H_{\rm Chen} }\right\|$ approaching to a constant as $t\to\infty$. 
In contrast, the Liouville model exhibits non-unitary evolution, 
where the evolution function diverges as $t$ increases.

\section{Extended models with PT-symmetric Liouville potentials}
\label{sec:ext}

Extended models of Liouville quantum mechanics can be categorized into those 
that maintain periodicity and those that disrupt it.
For instance, the model addressing spectral singularities in Ref.\ \cite{Correa:2012yq} preserves periodicity, 
while the model in Ref.\  \cite{Bender:2014bna} disrupts it.
In this section, we explore two typical models with Hamiltonians 
\begin{equation}
\label{eq:sum-exp}
H_c = p^2 +V_0\sum_{n=0}^\infty (-1)^n \me^{2 n \mi z},
\end{equation}
and 
\begin{equation}
\label{eq:bender}
H_b = p^2 -V_1 (\mi  z)^n \me^{2\mi z}+ V_2 \me^{2\mi z}.
\end{equation}
where $n$ is non-negative integers.

\subsection{Classical dynamics}

\begin{figure}[!ht]
     \centering
     \begin{subfigure}[b]{0.4\textwidth}
         \centering
         \includegraphics[width=\textwidth]{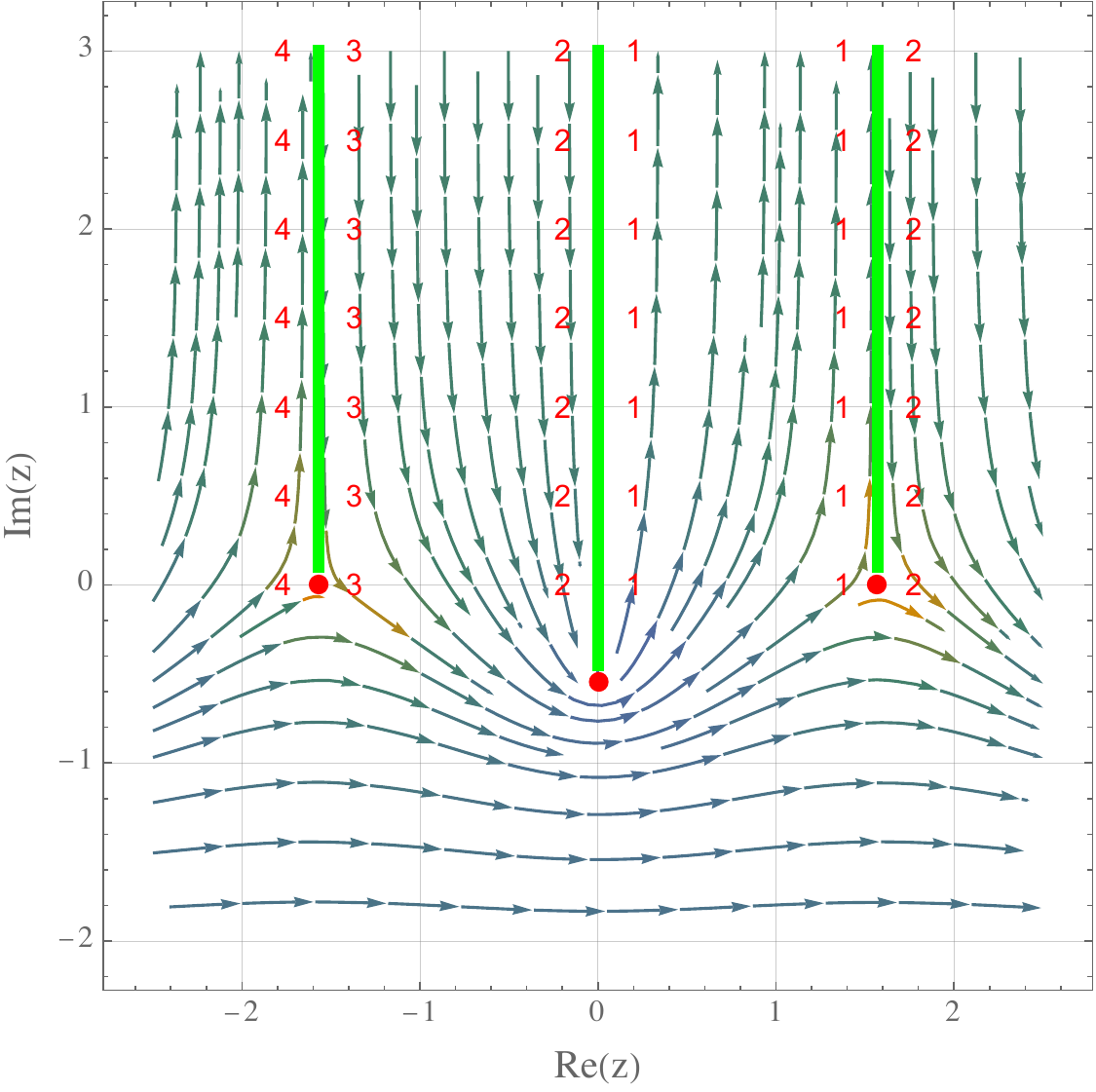}
         \caption{Upper sheet.}
         \label{fig:class-ext-exp2ix-1}
     \end{subfigure}
     \begin{subfigure}[b]{0.4\textwidth}
         \centering
         \includegraphics[width=\textwidth]{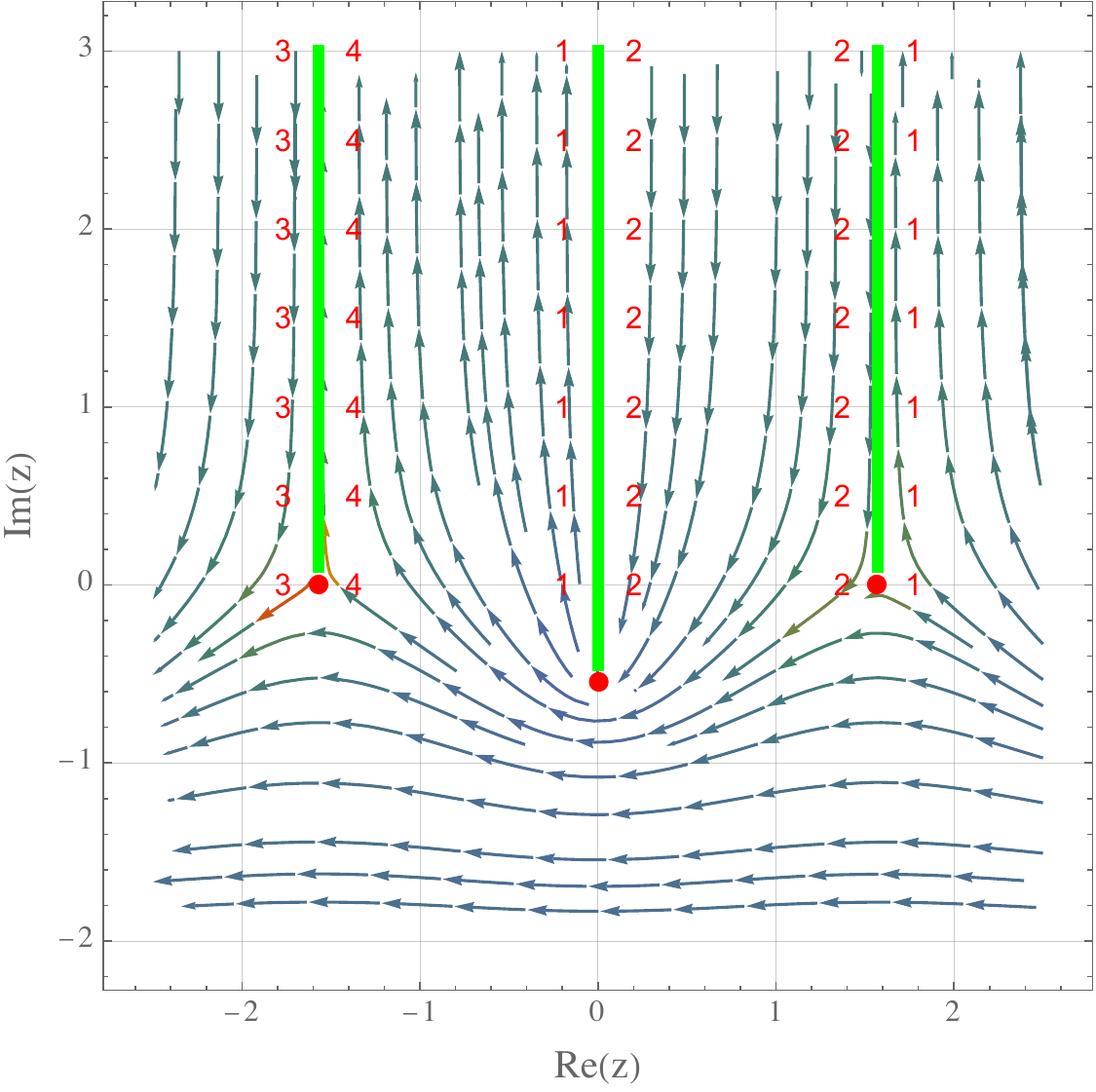}
         \caption{Lower sheet.}
         \label{fig:class-ext-exp2ix-2}
     \end{subfigure}
      \captionsetup{width=.9\textwidth}
       \caption{Phase portrait of $V=V_0/(1+\me^{2\mi z})$.}
        \label{fig:class-ext-exp2ix}
\end{figure}

First, we consider $H_c$, whose potential terms sum to 
\begin{equation}
V_c = \frac{V_0}{1+\me^{2\mi z}}
\end{equation}
This potential has periodic singularities $z_s$ and turning points $z_t$ ($\dot z =0$)
\begin{equation}
z_s= \frac{\pi }{2}+c_1 \pi ,\quad
z_t =\frac{1}{2\mi}\ln\left(\frac{V_0}{E}-1\right)+c_1 \pi ,\quad
c_1\in \mathbb{Z}.
\end{equation}
These points form all the branch points of the Riemann surface of the phase portrait. 
Each branch point has an index of $1/2$.
The branch cuts are chosen so that they all start from one point of $\{z_s\}$  and end at one point of $\{z_t\}$  
and belong to the set $\Im(V_c)=0$, see Fig.\ \ref{fig:class-ext-exp2ix}.
This implies that when creating a branch cut between $\pi/2$ and $\frac{1}{2\mi}\ln\left(\frac{V_0}{E}-1\right)$, we must first travel from $\pi/2$ to complex infinity, and then from complex infinity to $\frac{1}{2\mi}\ln\left(\frac{V_0}{E}-1\right)$. As a result, the point at complex infinity is inherently part of the branch cut.
Within the same sheet of a Riemann surface, edges that are denoted by the same number correspond to the same segment of a branch cut. Conversely, when the same number appears in different sheets of the Riemann surface, it indicates how those sheets are glued together.
Further, since no trajectories cross the branch cuts and no streams connect two fixed points, 
there are no closed orbits, which means that the spectrum is not discrete unless a compactification is introduced to the potential.

\begin{figure}[!ht]
     \centering
         \includegraphics[width=.5\textwidth]{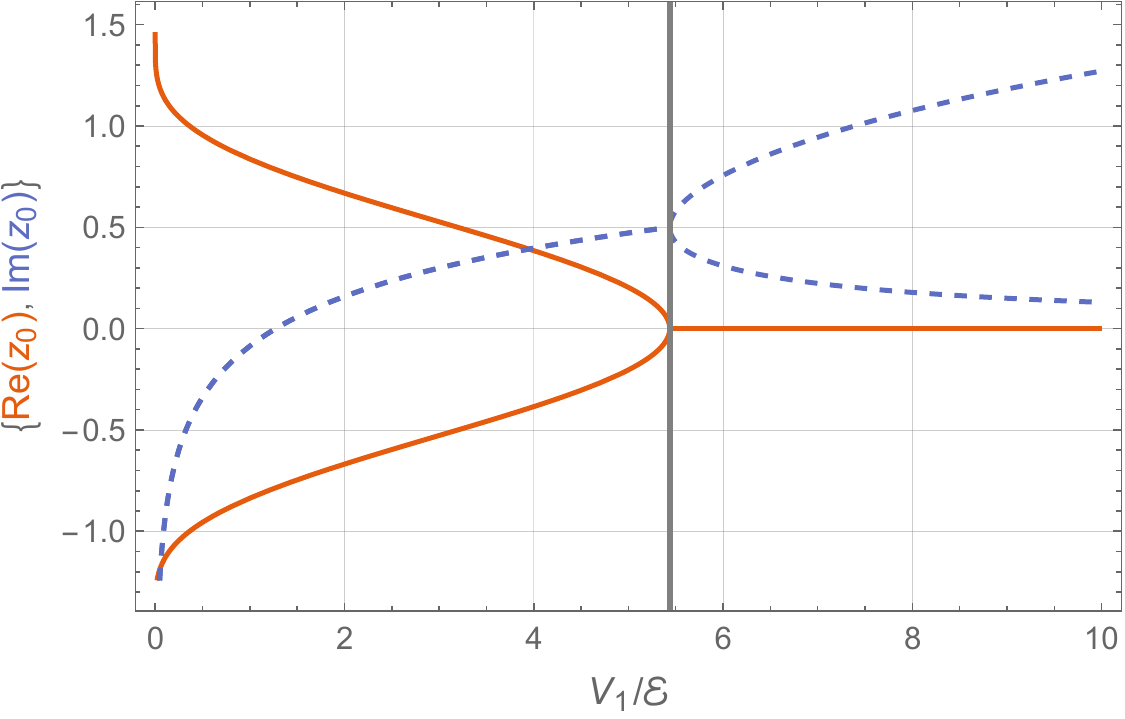}
      \captionsetup{width=.9\textwidth}
       \caption{Saddle-node bifurcation of central two branch points $V= -\mi V_1 z \me^{2\mi z}$}
        \label{fig:bifurcation}
\end{figure}

\begin{figure}[!ht]
     \centering
     \begin{subfigure}[b]{0.45\textwidth}
         \centering
         \includegraphics[width=\textwidth]{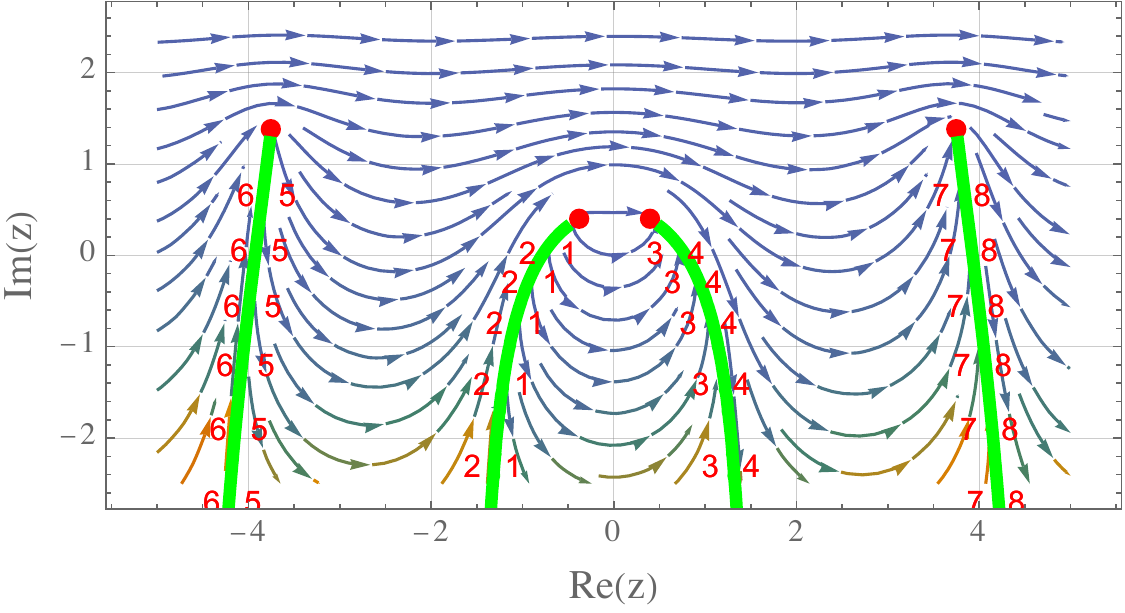}
         \caption{Upper sheet.}
         \label{fig:class-iznexp2iz-1}
     \end{subfigure}
     \begin{subfigure}[b]{0.45\textwidth}
         \centering
         \includegraphics[width=\textwidth]{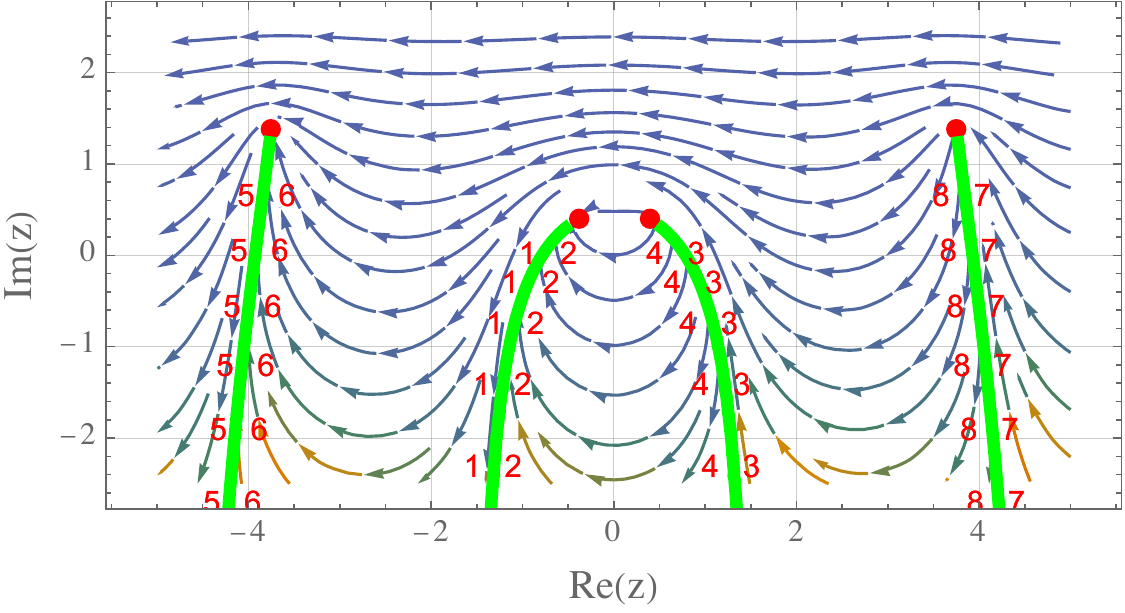}
         \caption{Lower sheet.}
         \label{fig:class-iznexp2iz-2}
     \end{subfigure}
      \captionsetup{width=.9\textwidth}
       \caption{$V= -\mi V_1 z \me^{2\mi z}$ with $V_1/E>2 \me$.}
        \label{fig:class-iznext-exp2ix}
\end{figure}

Now, let's examine the case of $H_b$ with $V_2=0$ and $n=1$. 
The branch points for this system are determined by equation $E+V_1 \mi z \me^{2\mi z}=0$, 
which can be expressed using the Lambert W function as
\begin{equation}
z=-\frac{\mi}{2}  W_{k}\!\left(-\frac{2 E}{V_1}\right),
\end{equation}
and each branch cut should originate from one of the points and be a member of the set $\{z| \Im(V)=0\}$.
This system is particularly complicated due to the behavior of the central two branch points
\begin{equation}
z_{-1}=-\frac{\mi}{2} W_{-1}\!\left(-\frac{2}{V_1/E}\right),\quad
z_0=-\frac{\mi}{2}  W_0\!\left(-\frac{2}{V_1/E}\right).
\end{equation}
These points undergo a saddle-node bifurcation as the parameter
 $V_1/E$ increases, as depicted in Fig.\ \ref{fig:bifurcation}.  
When $V_1/E<2 \me$, the central two branch points share the same imaginary part. 
The two-sheeted phase portrait is presented in Fig.\ \ref{fig:class-iznext-exp2ix}. 
It’s important to note that there are two distinct types of trajectories 
that can result in discrete spectra at the quantum level: those that connect 
$z_{-1}$ and $z_{0}$, and those that encircle these two branch points.

\begin{figure}[!ht]
     \centering
     \begin{subfigure}[b]{0.45\textwidth}
         \centering
         \includegraphics[width=\textwidth]{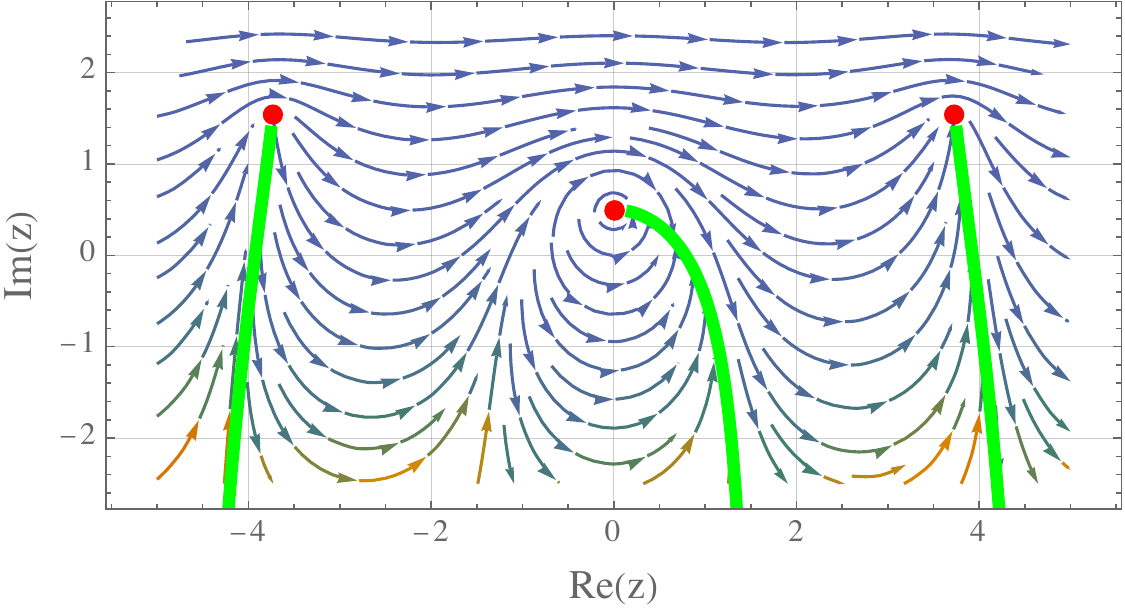}
         \caption{$V_1/E=2 \me$.}
         \label{fig:class-iznexp2iz-3}
     \end{subfigure}
     \begin{subfigure}[b]{0.45\textwidth}
         \centering
         \includegraphics[width=\textwidth]{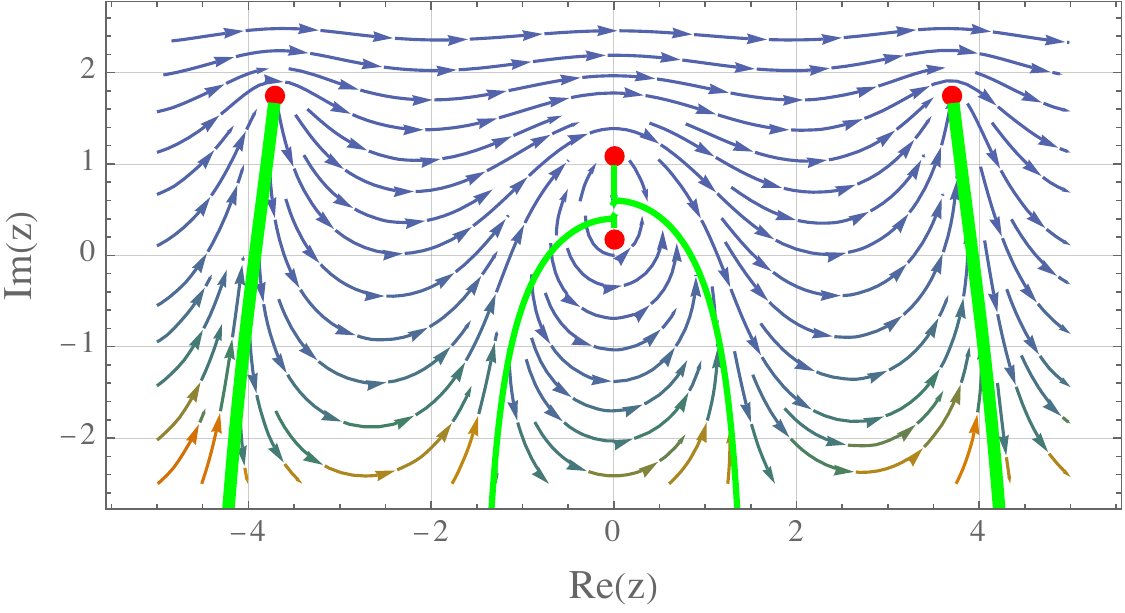}
         \caption{$V_1/E<2 \me$.}
         \label{fig:class-iznexp2iz-4}
     \end{subfigure}
      \captionsetup{width=.9\textwidth}
       \caption{Alternative cases for $V= -\mi V_1 z \me^{2\mi z}$.}
        \label{fig:class-ext-exp2ix-sum}
\end{figure}

When $V_1/E=2 \me$, the two distinct points merge into one, 
and two branch cuts are reduced to a single cut. 
For instance, we can choose to retain the branch cut where the streamlines converge 
while removing the one from which the streamlines emit, refer to the left panel in Fig.\ \ref{fig:class-iznexp2iz-3}; 
conversely, the inverse action is also permissible. 
In terms of the Riemann surface, the streamlines are incapable of forming a closed trajectory. 
This is because even though the branch cut of streamline emission is removed, 
a barrier to the streamline progression remains at the original location. 
That is, on either side of this barrier, the streamlines flow in opposite directions. 
Specifically, a streamline initiating from one side of the barrier will initially move towards the branch cut, 
cross over to the second sheet of the Riemann surface, 
and eventually, return to this barrier on the latter sheet, 
where it ceases its movement. This does not constitute a closed trajectory in the sense of the traditional dynamical system.

When  $V_1/E>2 \me$, the two points have the same real parts.
The upper sheet is shown in Fig.\ \ref{fig:class-iznexp2iz-4}.
The branch cuts extending from the two central points are more intricate than those in the previous examples. 
We propose a method for drawing these cuts: 
one branch cut extends down the imaginary axis from the upper branch point, 
then bends towards the cut where the streamlines converge on the right; 
another branch cut rises the imaginary axis from the lower branch point before turning towards the left cut 
where the streamlines emit on the left.
On such a Riemann surface, there exists a class of closed orbits that encircle two central branch points and bouncing trajectories 
that connect these two branch points.


\subsection{Equivalent Hermitian Hamiltonian}


To apply the supershift used for a single Liouville potential, given by the transformation
\begin{equation}
\label{eq:transtoinf}
z\to z+\frac{\mi}{2} \arctanh(\e),
\end{equation}
we first decompose the Hamiltonian 
\begin{equation}
H_c = p^2 +\frac{V_0}{1+\me^{2 \mi z}}
\end{equation}
into its Hermitian and anti-Hermitian parts and insert the parameter $\e$. This yields
\begin{equation}
H_c=p^2 +V_0\frac{1+\cos 2z-\mi \e  \sin 2z}{\sin ^2 2z +(1+\cos  2z)^2},
\end{equation}
After applying the shift in Eq.\ \eqref{eq:transtoinf}, we obtain
\begin{equation}
h_c^\e=p^2 +V_0 \frac{\sqrt{1-\e ^2}+\left(1+\e^2\right) \cos 2z-2 \mi \e  \sin 2z}{2 \left(\sqrt{1-\e ^2}+\cos 2z-\mi \e  \sin 2z\right)},
\end{equation}
where the momentum operator $p$ remains unchanged. 
Taking the limit, we find  the Hermitian equivalence of $H_c$
\begin{equation}
\lim_{\e\to1} h_c^\e=p^2+V_0.
\end{equation}
Since $V_0$ is a constant, it can be absorbed into the ground-state energy, 
and so we get that $H_c$ is also equivalent to the Hamiltonian of a free particle.
Since $V_0$ is a constant, it can be incorporated into the ground-state energy. 
As a result, we find that $H_c$ is also equivalent to the Hamiltonian of a free particle, 
which is simply the kinetic energy term $p^2$.

To analyze the Hamiltonian $H_b$, we categorize it into two distinct cases based on whether the powers of $n$  are even or odd
\begin{subequations}
\begin{equation}
H_b^{\rm odd} = p^2 -\mi V_1 (-1)^{k}  z^{2k+1} \me^{2\mi z}+ V_2 \me^{2\mi z},
\end{equation}
\begin{equation}
H_b^{\rm even} = p^2 -V_1 (-1)^{k}  z^{2k} \me^{2\mi z}+ V_2 \me^{2\mi z},
\end{equation}
\end{subequations}
where $k\in\mathbb{Z}$.
We then apply the same transformation procedure as before,  substituting $z\to z+\frac{\mi}{2} \arctanh(\e)$, 
which gives us the following expressions
\begin{subequations}
\begin{equation}
h_b^{\e,\, {\rm odd}} =
p^2+V_2 \sqrt{1-\e ^2} \cos 2z+
(-1)^k V_1\sqrt{1-\e ^2} \left(z+\frac{\mi}{2} \arctanh \e \right)^{2 k+1} \sin 2z,
\end{equation}
\begin{equation}
h_b^{\e,\, {\rm even}} =p^2+V_2\sqrt{1-\e ^2} \cos 2x 
-(-1)^k V_1 \sqrt{1-\e ^2} \left(z+\frac{\mi}{2} \arctanh \e \right)^{2 k}\cos 2 z.
\end{equation}
\end{subequations}
As $\e\to 1$, the terms $\sqrt{1-\e ^2}\arctanh^n \e$ tends towards zero for finite $n$.
Consequently, in the limit $\e\to 1$, both $h_b^{\e,\, {\rm odd}}$ and $h_b^{\e,\, {\rm even}} $ 
reduce to the simple form of the Hamiltonian of a free particle.
\begin{equation}
\lim_{\e\to 1} h_b^{\e,\, {\rm odd}} =\lim_{\e\to 1} h_b^{\e,\, {\rm even}}  = p^2.
\end{equation}

The result for $H_b$ is somewhat surprising because it possesses a natural closed orbital, 
whereas $H_c$ and Liouville quantum mechanics do not have closed orbitals. 
This distinction suggests that $H_b$ would exhibit a discrete energy spectrum at the quantum level, 
while for $H_c$ and Liouville quantum mechanics, 
one would need to impose periodic boundary conditions artificially to achieve a discrete spectrum. 
This implies that the transformation used may discard certain information, 
such as the specifics of the periodic boundary conditions, 
thereby altering the fundamental nature of the system's dynamics.

\subsection{Pseudospectrum analysis}

\begin{figure}[!ht]
     \centering
         \includegraphics[width=.8\textwidth]{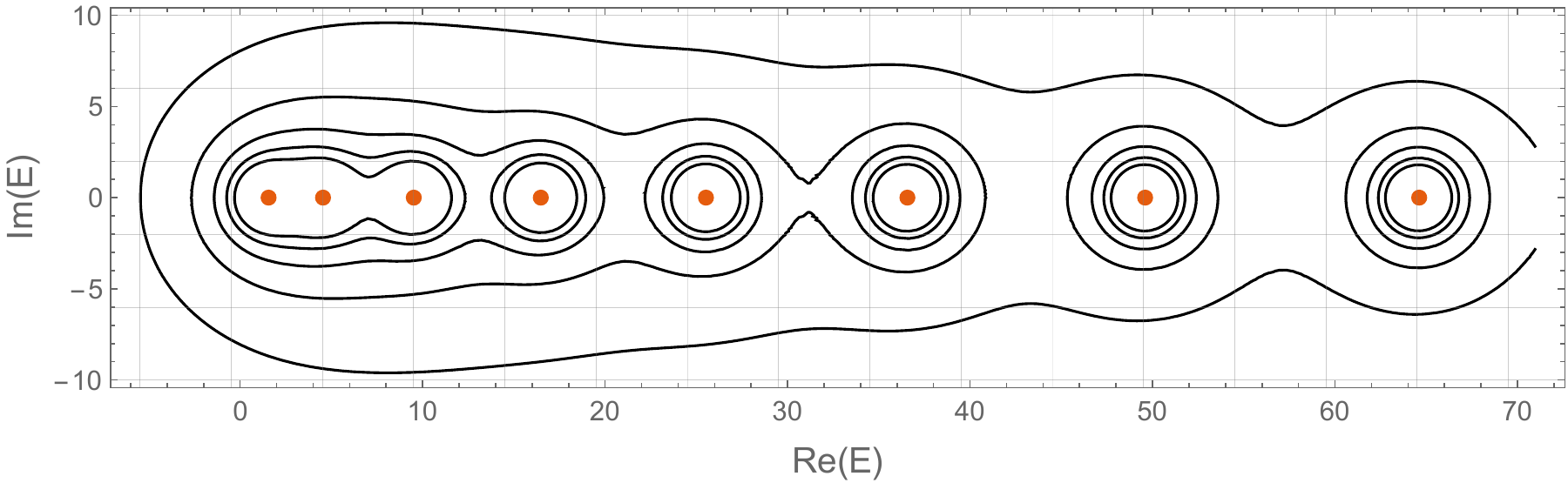}
      \captionsetup{width=.9\textwidth}
       \caption{$V=V_0/(1+\me^{2\mi z})$}
        \label{fig:ext-exp2ix}
\end{figure}

The compactification parameter is set at $L=\pi/2$, and the energy spectra of $H_c$ are presented 
in Tab.\ \ref{tab:eg-exp}.
Although the imaginary part of the energy spectrum is non-zero, 
it is relatively small compared to the real part, 
likely due to the precision of the computational method.
\begin{table}[!ht]
\begin{center}
\begin{tabular}{c|cccc}
\hline
$n$ & 1 & 2 & 3 & 4 \\
\hline
$E_n$ & $1.563 -0.001 \mi$ & $4.520 -0.002 \mi$ & $9.518-0.003 \mi$ & $16.524 -0.004 \mi$ \\
\hline
$n$ & 5 & 6 & 7 & 8 \\
\hline
$E_n$ & $25.534 - 0.005 \mi$ & $36.547 - 0.006 \mi$ & $49.563 - 0.007 \mi$ & $64.582 - 0.008 \mi$ \\
\hline
$n$ & 9 & 10 & 11 & 12 \\
\hline
$E_n$ & $81.604 - 0.009 \mi$ & $100.628 - 0.009 \mi$ & $121.654 - 0.010 \mi$ & $144.683 - 0.011 \mi$ \\
\hline
\end{tabular}
\end{center}
\captionsetup{width=.9\textwidth}
\caption{Eigenvalues of $H_c$.\label{tab:eg-exp}}
\end{table}%
By fitting the real part of the first $37$ energy levels, we obtain the expression $E_n = c (n+ d)^2$
with $c = 1.00008$ and $d= 0.0267178$.
This result is very close to that of the Liouville model,
as both are effectively equivalent to the Hermitian Hamiltonian $H=p^2$.
The pseudospectrum is shown in Fig.\ \ref{fig:ext-exp2ix}.

\begin{figure}[!ht]
     \centering
     \begin{subfigure}[b]{0.45\textwidth}
         \centering
         \includegraphics[width=\textwidth]{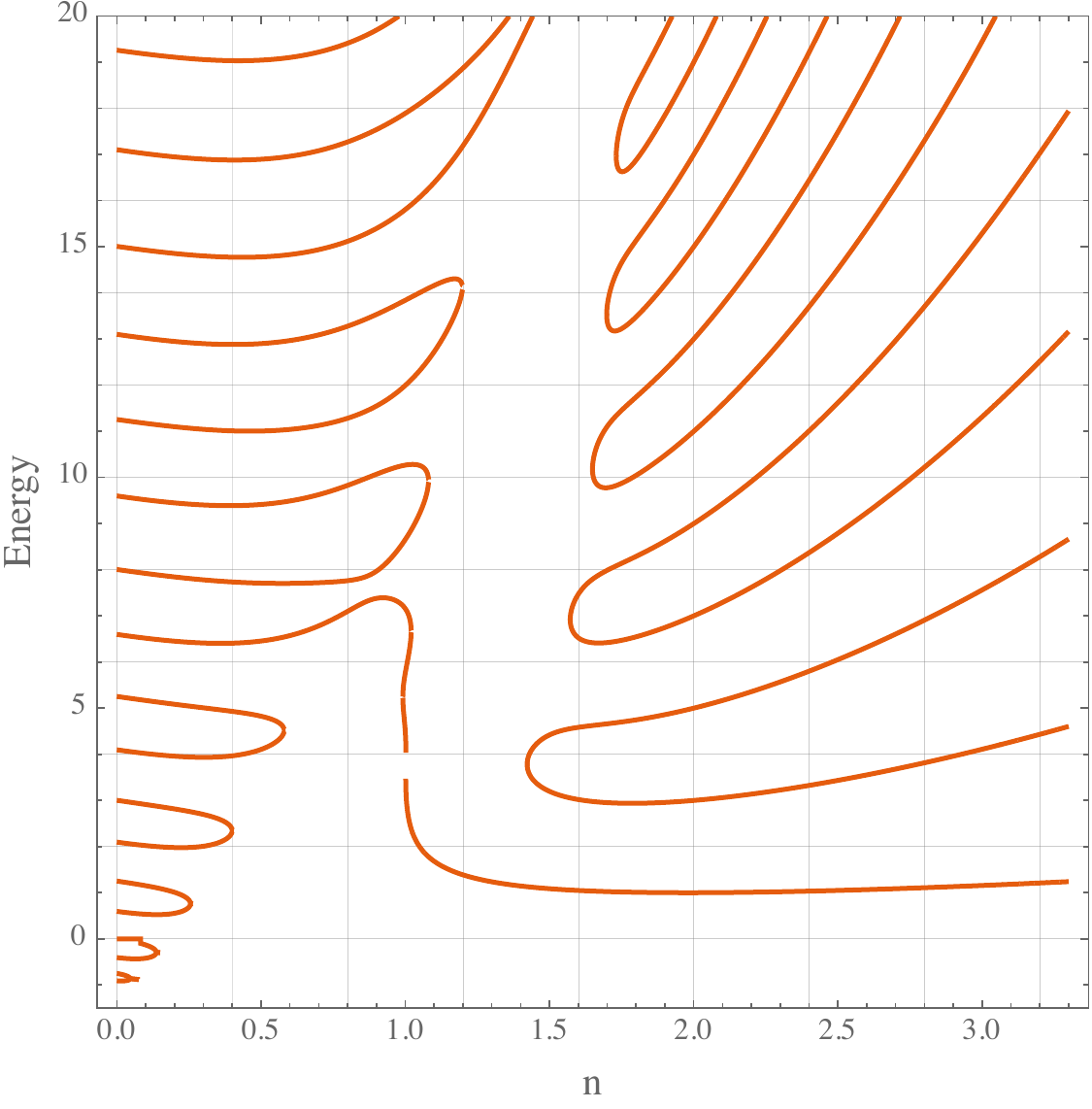}
         \caption{$V=-(\mi z)^n$.}
         \label{fig:izn}
     \end{subfigure}
     \begin{subfigure}[b]{0.45\textwidth}
         \centering
         \includegraphics[width=\textwidth]{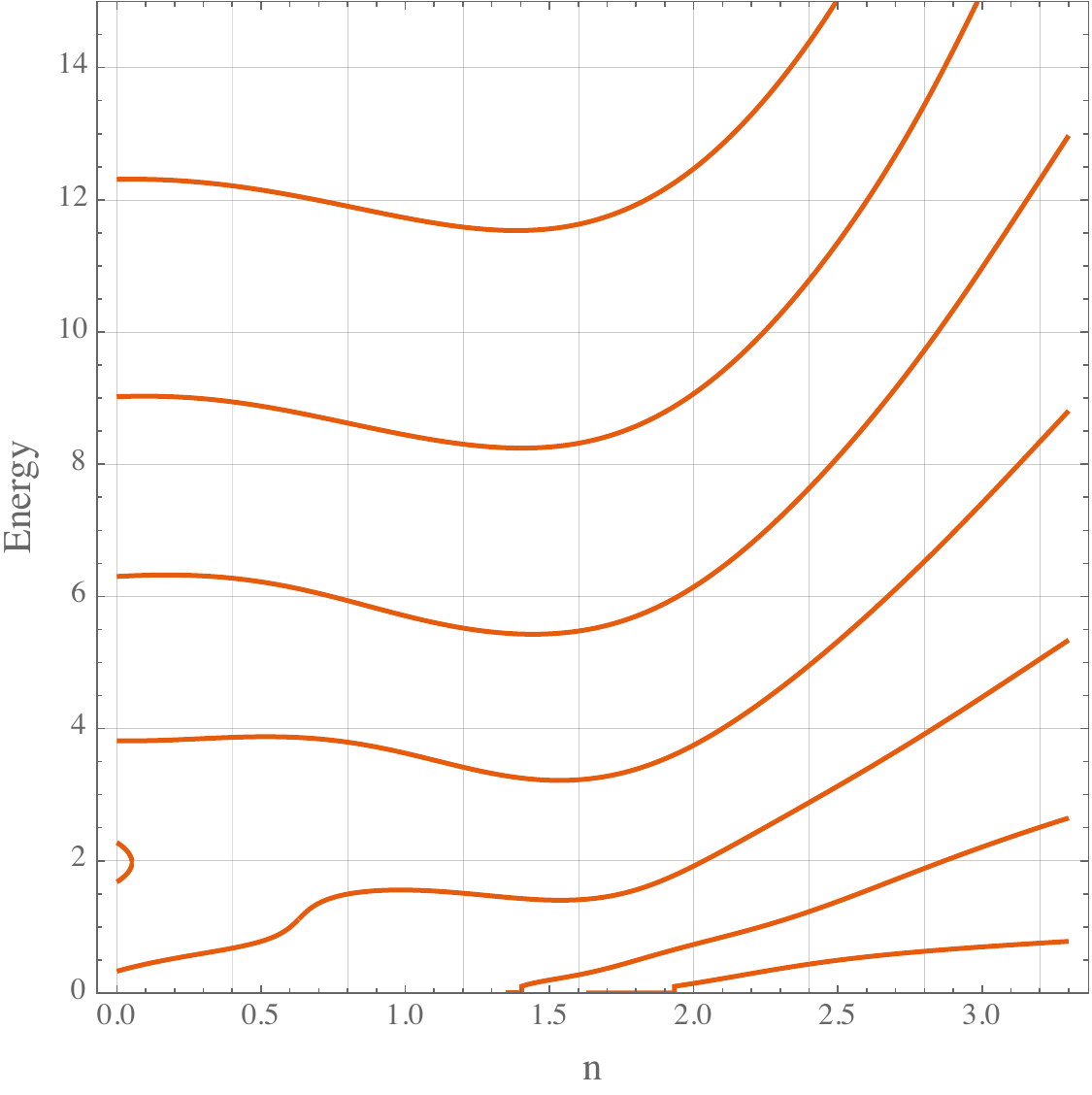}
         \caption{$V=-(\mi z)^n\me^{2\mi z}$.}
         \label{fig:iznexp2iz}
     \end{subfigure}
      \captionsetup{width=.9\textwidth}
       \caption{Comparison of two models with respect to the parameter $n$.}
        \label{fig:izn_exp2iz}
\end{figure}

For the model $H_b$ with $V_1=1$ and $V_2=0$, we compared it to $V=-(\mi z)^n$. 
Fig.\ \ref{fig:izn} illustrates the spectra of $V=-(\mi z)^n$ with respect to $n$, 
obtained using the spectral method.
The compactification parameter is $L=2\pi$  
because the wave function becomes negligible beyond this interval.
Fig.\ \ref{fig:izn} reveals the same pattern for $n>1$ as reported in Ref.\ \cite{Bender:1998ke}.
The break in the data curve around $n=1$ corresponds to complex values.
In the region $1>n>0$,  the spectra method provides an additional pattern, and when $n=0$, 
the spectrum matches that of an infinite potential well of width $2L$.

Fig.\ \ref{fig:iznexp2iz} illustrates the spectral nature of the $H_b$ model with parameters $V_1=1$ and $V_2=0$, 
given the compactification parameter $L=\pi$ due to the periodic boundary condition. 
The introduction of the complex Liouville potential dramatically changes the spectral characteristics. 

\begin{figure}[!ht]
     \centering
         \includegraphics[width=.9\textwidth]{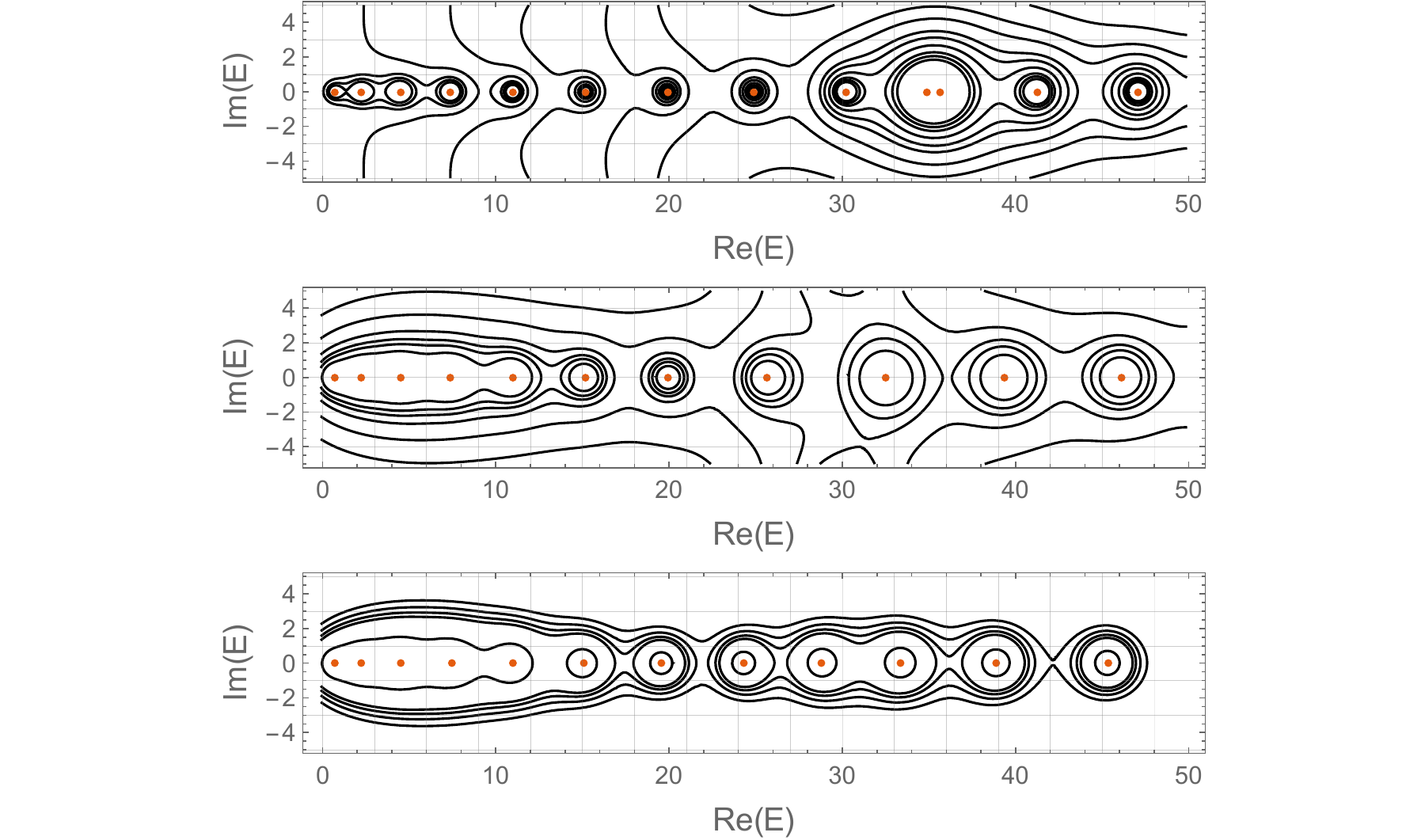}
      \captionsetup{width=.9\textwidth}
       \caption{Variation in spectral instability for the potential $V=-  (\mi z)^3 \me^{2\mi z}$, with compactification parameters (from top to bottom) $L=2\pi,\, \pi,\, \pi-0.2$.}
        \label{fig:instability}
\end{figure}

Moreover, the $H_b$ model with $n=3$ exhibits the spectral instability. 
A similar instability phenomenon is observed in Bender and Boettcher's model \cite{Bender:1998ke} at $n=3$ \cite{Krejcirik:2014kaa}. 
Our findings suggest that the nature of spectral instability evolves as the compactification parameter is diminished. 
A critical transition occurs at $L_c=\pi$, as depicted in Fig.\ \ref{fig:instability}.
The sequence of plots, from top to bottom, begins with the pseudospectrum at $L=2\pi$, 
where open contours signal instability. The middle plot, with the critical value $L_c=\pi$, 
demonstrates a marked improvement in stability. 
The bottom plot, where $L=\pi-0.2$, reveals a complete transformation of the instability nature as the compactification parameter 
crosses the critical threshold $L_c$.

\section{Conclusions}
\label{sec:conclusion}

In this study, we delved into the PT-symmetric Liouville theory and its variants, 
employing a reformed index theory for classical examination and the SCM for quantum analysis.

For the index theory, we extended conventional phase-portrait techniques \cite{Strogatz:2024ndc} 
to incorporate two-sheeted Riemann surfaces, providing a more intuitive framework 
compared to the traditional approach \cite{Bender:2023cem,Curtright:2005zk}, 
which treats PT-symmetric systems as two-dimensional real mechanical models 
and disregards the multivalued nature of complex functions. 
This extension enables us to determine the existence of closed orbits in dynamical systems on Riemann surfaces. 
Unlike the traditional theory \cite{Strogatz:2024ndc}, where the index theorem applies only to real systems, 
our approach accommodates complex systems with fractional fixed-point indices, 
providing a more comprehensive understanding of the underlying dynamics.


For the SCM, we observed an interesting phenomenon: 
even when the differential Hamiltonian operator preserves Hermiticity, 
the corresponding Hamiltonian matrix formed from the CDM does not. 
Consequently, the resulting eigenstates lack completeness. 
Expanding on this discovery, we revisited the Liouville model and its variations, 
uncovering the restoration of the eigenstates’ completeness. 
This result is achieved by leveraging SCM's ability to transform differential operators into matrices, 
simplifying the process of finding similarity matrices that map the non-Hermitian theory to its Hermitian equivalent \cite{Mostafazadeh:2003gz}.
Furthermore, we observed that the pseudospectra’s stability depends on the compactification parameter.
This suggests that in PT-symmetric theories, additional information may be required when using SCM to analyze spectral stability, owing to the periodic or multivalued nature of the complex potential.

The non-Hermitian character of Chebyshev differential matrices raises several questions, 
such as
the appearance of exceptional points in the transition from Hermitian to non-Hermitian matrix representations of differential operators,
and why the corresponding Hermitian differential operator shares the same spectrum. 
These issues will be investigated further in our ongoing research.

\section*{Acknowledgement}

L.C. and H.G. are supported by Yantai University under Grant No.\ WL22B224 and No.\ WL24B06.

\appendix

\section{Perturbative calculation for metric operator}
\label{app:pert}

A Hamiltonian operator $H$ is termed pseudo-Hermitian if there exists a Hermitian operator $\eta = e^Q$ such that $H= \eta^{-1} H^\dag \eta$, where $\eta$ is known as the metric operator \cite{Mostafazadeh:2008pw}. Moreover, if $\eta$ is positive definite, one can forge a Hermitian Hamiltonian $h=\rho H\rho^{-1}$ using its Hermitian square root $\rho=e^{Q/2}$, while the new theory with Hamiltonian $h$ retains the same energy spectrum as the original theory.
Hence, the pivotal task in calculating the Hermitian equivalent Hamiltonian is to determine the operator $Q$. One approach to achieve this is through perturbation theory, which begins by dividing the initial Hamiltonian into two components:
\begin{equation}
H^\e=H_0 +\e H_1
\end{equation}
Here, $H_0$ is Hermitian and $H_1$ is anti-Hermitian, meaning $H^\dag_0 = H_0$ and $H^\dag_1 =- H_1$. The parameter $\e$ is introduced to facilitate perturbation calculations. Consequently, $Q$ and $h$ can be expressed as series expansions in $\e$:
\begin{equation}
Q=\sum_{\rm n, odd} Q_n \e^n,\quad
h=\sum_{\rm n, even} h_n \e^n
\end{equation}
Using the definition of pseudo-hermiticity, we obtain
\begin{equation}
H_0-\e H_1 = e^{Q} H_0 e^{-Q} + 
		\e e^{Q} H_1 e^{-Q}
\end{equation}
The expression $e^Q H e^{-Q}$ can be determined by an infinite sequence of commutators:
\begin{equation}
e^Q H e^{-Q} = H +[Q, H]+\frac{1}{2!}[Q,[Q,H]]+
\frac{1}{3!}[Q,[Q,[Q,H]]]+\ldots
\end{equation}
After substituting the expansions for $Q$ and $H$, the first few equations for the coefficients $Q_n$ are derived:
\begin{equation}
\label{eq:pert-Q}
\begin{split}
\e^1:\qquad  [H_0, Q_1] = &2 H_1 \\
\e^3:\qquad  [H_0, Q_3] =& -\frac{1}{6} [Q_1,[H_1,Q_1]]\\
\e^5:\qquad  [H_0, Q_5] = & -\frac{1}{6} \left( [Q_1,[H_1,Q_3]]+ [Q_3,[H_1,Q_1]]\right)+\\
&\qquad +\frac{1}{360}[Q_1,[Q_1, [Q_1, [H_1, Q_1]]]]
\end{split}
\end{equation}
Utilizing these equations, we can compute the initial terms of $h= e^{Q/2} H e^{-Q/2}$:
\begin{equation}\begin{split}
\e^0:\qquad  h_0 = & H_0 \\
\e^2:\qquad  h_2 =& -\frac{1}{4} [H_1,Q_1]\\
\e^4:\qquad  h_4 = & -\frac{1}{4} [H_1, Q_3]+
\frac{1}{192} [Q_1, [Q_1, [H_1, Q_1]]]
\end{split}\end{equation}
Typically, $h$ does not have a closed form and may diverge as $\e$ tends towards unity.

\bibliographystyle{utphys}
\bibliography{references}
\end{document}